\documentclass[article,twocolumn,aps,superscriptaddress,longbibliography]{revtex4-2}

\usepackage{graphicx}
\usepackage{dcolumn}% Align table columns on decimal point
\usepackage{bm}% bold math
\usepackage{epstopdf}
\usepackage{float}
\usepackage[colorlinks=true, linkcolor=blue, citecolor=blue, urlcolor=blue]{hyperref}
\usepackage{amsmath}
\usepackage{xcolor}
\usepackage{comment}
\usepackage[normalem]{ulem}
\usepackage{physics}
\graphicspath{{./images/}}

\begin{document}
	\title{Topological quantum compilation of metaplectic anyons based on the genetic optimized algorithms}
	
	\author{Jiangwei Long}
	\affiliation{School of Physics and Optoelectronics, Xiangtan University, Xiangtan 411105, Hunan, People’s Republic of China.}
	\author{Jianxin Zhong}
	\email{jxzhong@shu.edu.cn}
	\affiliation{Center for Quantum Science and Technology, Department of Physics, Shanghai University, Shanghai 200444, People’s Republic of China.}
	\affiliation{Hunan Key Laboratory for Micro-Nano Energy Materials and Devices, Hunan, People’s Republic of China.}
	\author{Lijun Meng}
	\email{ljmeng@xtu.edu.cn}
	\affiliation{School of Physics and Optoelectronics, Xiangtan University, Xiangtan 411105, Hunan, People’s Republic of China.}
	\affiliation{Hunan Key Laboratory for Micro-Nano Energy Materials and Devices, Hunan, People’s Republic of China.}
	
	\date{\today}

	\begin{abstract} 
		Topological quantum computing holding global anti-interference ability is realized by braiding some anyons, such as well-known Fibonacci anyons. Here, based on $SO(3)_2 $ theory we obtain a total of 6 anyon models utilizing \textit{F}-matrices, \textit{R}-symbols, and fusion rules of metaplectic anyon.We obtain the elementary braiding matrices (EBMs) by means of unconventional encoding. After braiding \textit{X} and $X^\prime$, we insert a pair of \textit{Z} anyons into them to ensure that the initial order of anyons remains unchanged. In this process only fusion is required, and measurement is not necessary. Three of them $\{V^{113}_3,V^{131}_3,V^{133}_1\}$ are studied in detail. We study systematically the compilation of these three models through EBMs obtained analytically. For one-qubit case, the classical \textit{H}- and \textit{T}-gate can be well constructed using the genetic algorithm enhanced Solovay-Kitaev algorithm (GA-enhanced SKA) by $\{V^{113}_3,V^{131}_3,V^{133}_1\}$. The obtained accuracy of the \textit{H}/\textit{T}-gate by $\{V^{113}_3,V^{133}_1\}$ is slightly inferior to the corresponding gates of the Fibonacci anyon model, but it also can meet the requirements of fault-tolerant quantum computing, $V^{131}_3$ giving the best performance of these four models. For the two-qubit case, we use the exhaustive method for short lengths and the GA for long lengths to obtain braidword for $\{V^{113}_3,V^{131}_3,V^{133}_1\}$ models. The resulting matrices can well approximate the local equivalence class of the CNOT-gate, while demonstrating a much smaller error than the Fibonacci model, especially for the $V^{113}_3$.The braiding processes of conventional encoding (using identical anyons) and unconventional encoding (using distinct anyons) are compared. Finally, we attempt to generalize the model to the \textit{N}-qubit case.  
	\end{abstract}
	
	\maketitle
	\section{Introduction}
	The anyon is a quasi-particle excitation that exists in a two-dimensional plane, and the exchange of two anyons will result in a nontrivial phase, which was first proposed by Myrheim and Leinaas \cite{RN1} in 1977. In 1997, Kitaev proposed that the use of anyons is expected to realize topological quantum computing for the first time \cite{RN2}. The advantage of topological quantum computing is that information is stored globally and immune to local noise, making it naturally robust. Topological quantum computing relies on the braiding of non-Abelian anyons \cite{RN3,RN4,RN5,RN53,RN54}. Notable non-Abelian anyons include the Ising and Fibonacci anyons. Ising anyon can easily construct \textit{H}-gate and CNOT-gate but cannot build \textit{T}-gate \cite{RN6,RN7,RN8,RN9,RN10}. Non-Abelian Fibonacci anyon can be used for universal quantum computing by braiding alone and has been discussed extensively in the past decades \cite{RN11}. The standard \textit{H}-/\textit{T}-gate cannot be directly implemented with Fibonacci anyons, requiring continuous braid operations to reduce errors to a level comparable to standard qubit gates. This sequence is called a braidword, which consists of some elementary braiding matrices (EBMs). As the length of the braidword increases, the search space will increase exponentially. Exhaustive search will quickly become infeasible due to high computational cost. How to find the best approximation between braidword and standard gate in an exponentially large space forms the quantum compilation problem. Many methods have been proposed for the quantum compilation problem of the Fibonacci anyon model. Including hash function techniques \cite{RN12}, SKA \cite{RN13}, Monte Carlo enhanced SKA \cite{RN14}, machine learning \cite{RN15}, and GA \cite{RN16}. For the construction of two-qubit gate using Fibonacci anyon, L. Hormozi et al. proposed the injection braid method \cite{RN17}, Haitan Xu proposed functional braid to construct CNOT-gates with low leakage error \cite{RN18}, two-qubit EBMs based on Fibonacci model are given by Cui et al. \cite{RN19} and Phillip C. Burke et al. recently used these EBMs to construct a local equivalence class with CNOT-gate \cite{RN20}. Haitan Xu et al. mapped between dense and sparse representations of qubits in anyon models supported by $ SU(2)_k$ theory to construct multi-qubit gates, they first proposed the construction of a three-qubit gate using Fibonacci anyons\cite{RN21}. Fibonacci anyon was also used to construct a three-qubit gate, which was theoretically realized by Abdellah Tounsi et al. using an injection braid \cite{RN22}.
	
	Although Fibonacci anyon can be used to realize universal quantum computing by braid alone, such anyon seems to be difficult to capture experimentally. A more easily implementable class of anyons for universal quantum computing is called weak integral anyon (corresponding to the quantum dimension $d^2\in Z$) \cite{RN23}, and weak integral anyon with \textit{F}-property \cite{RN24} includes metaplectic anyons \cite{RN25,RN26,RN27}. Metaplectic anyons model $SO(p)_2$, when p=3, is equivalent to $SU(2)_4$, the anyons in the $SO(3)_2$ are expected to exist in fractional quantum Hall liquids with $v = 8/3$ \cite{RN28}, bilayer fractional quantum Hall liquids with $v = 2/3$ \cite{RN29} and parafermion particle zero modes \cite{RN30}.Recent advancements in generating fractional quantum Hall states with ultracold atoms \cite{RN50}, combined with the experimental observation of the fractional anomalous quantum Hall effect in twisted bilayer molybdenum ditelluride ($MoTe_2$)  \cite{RN51} and orthorhombic five-layer graphene systems \cite{RN52}, have opened new avenues for realizing metaplectic anyons."
	
	In the metaplectic anyons system $SO(p)_2$, the simple objects (anyon types) are denoted as $\{1,X_\epsilon,Y_j,X^\prime _\epsilon ,Z,1\leq j\leq r\}$, where $p=2r+1$. The case of \textit{p}=3 is the center of our interest, in this case $SO(3)_2 = SU(2)_4 $, the simple objects become $\{1,X_\epsilon,Y,X^\prime _\epsilon ,Z\}$ and their corresponding topological spins are \{0,1/2,1,3/2,2\} , following reference \cite{RN27}, which labeled them twice topological spins \{0,1,2,3,4\}. For simplicity, we use $X/X^\prime$ for $X_\epsilon/X_\epsilon^\prime $ later.
	
	According to the k-level theory of fusion rules \cite{RN31}, the topological spins $s_1 $ and $s_2$ can be fused (recoupling,$\otimes$) and give total coupling spins 
	
	\begin{equation}
		\large
		\begin{split}
			s_1\otimes s_2 = |s_1-s_2|\oplus|s_1-s_2|+1...\\...\oplus min(s_1+s_2, k-s_1-s_2),
		\end{split}
		\label{eq.1}
	\end{equation}
	
	where $\otimes$ and $\oplus$ denote direct product and direct sum respectively. In Eq .\eqref{eq.1} combined with the topological spins of the metaplectic anyons, we can get all the fusion rules that need to be used:
	
	\begin{equation}
		\small
		\begin{split}
			X\otimes X=1\oplus Y,X\otimes Y=X\oplus X^\prime,
			X\otimes X^\prime=Y\oplus Z,\\
			X\otimes Z=X^\prime,Y\otimes Y=1\oplus Y\oplus Z,
			Y\otimes X^\prime=X\oplus X^\prime,\\
			Y\otimes Z=Y,X^\prime\otimes X^\prime=1\oplus Y,
			X^\prime\otimes Z=X,Z\otimes Z=1.
		\end{split}
		\label{eq.2}
	\end{equation}

	Note that the fusion rules of anyons and vacuum are not presented. Based on the above fusion rules, combining several \textit{F}-matrices and \textit{R}-symbols of metaplectic anyons,  EBMs for constructing standard quantum gates can be obtained, including one-qubit and two-qubit gates. The methodology for deriving EBMs is predicated on established solution protocols previously developed for both Fibonacci anyon and metaplectic anyon models \cite{RN27,RN48}. The standard set of universal gates for qubit quantum circuit models consists of a Hadamard gate \textit{H}, a phase gate \textit{T}, and a controlled-not gate CNOT \cite{RN32,RN33}. Therefore, it is important to construct the set of those quantum gates using EBMs and measure the distance of a braidword from a standard qubit gate as the precision index of the constructed quantum gate.
	
	Cui and Wang demonstrated \cite{RN27} that the combination of the braidings of the anyon \textit{X} and the projective measurement of the total charge of the two metaplectic anyons can implement universal quantum computing. Previous studies identified the universality of a qutrit model $V^{1111}_2$ and a qubit model $V^{1111}_0$ \cite{RN27}, $V^{2222}_0$ and $V^{1221}_0$ models \cite{RN34}. For constructing a two-qubit gate, we can choose to use 8 anyons (4 anyons corresponding to one qubit) \cite{RN22,RN27,RN35} or 6 anyons (3 anyons corresponding to one qubit) \cite{RN17,RN21,RN36}. Choosing the latter (6 anyons) to construct a two-qubit gate because the number of redundant non-computational states can be reduced effectively. As a consequence, the dimension of EBMs can be kept at a low-rank level, which is conducive to the solution of EBMs of a two-qubit gate. Standard encoding uses three or four \textit{X} with a topological spin of 1/2, and universal quantum computing cannot be implemented under this limitation in $SO(3)_2 $. But it has been shown that braiding with fusion and measurement of projection  together can achieve a universal set of gates \cite{RN37}. By introducing a pair of \textit{Z} anyons after braiding \textit{X} and $X^\prime$ once, the arrangement of initial state anyons can be recoverd without affecting the intermediate state, which makes non-standard coding feasible. In this process, fusion is necessary, but measurement does not seem to be necessary. There are 6 qubit models in the metaplectic anyon model with 3 anyons encoding, among which 3 models $\{V^{113}_3,V^{131}_3,V^{133}_1\}$ are selected for the systematic construction of one/two-qubit gate, refer to Appendix A for a detailed analysis procedure.
	
	The structure of this article is as follows. In the part \uppercase\expandafter{\romannumeral2}, the one-qubit EBMs of $\{V^{113}_3,V^{131}_3,V^{133}_1\}$ is given, and show the results of constructing \textit{H}- and \textit{T}-gates which are optimized by GA-enhanced SKA. In the part \uppercase\expandafter{\romannumeral3},  the EBMs of  $\{V^{113}_3,V^{131}_3,V^{133}_1\}$ is presented for constructing two-qubit gates, and show the approximation of braidword to a local equivalence class of CNOT-gates. In the part \uppercase\expandafter{\romannumeral4}, we summarize our results. Appendix A analyzes the process of selecting the optimal qubit model among all metaplectic anyon models with 3 anyons encoding. Appendix B lists all the basis transformation matrices \textit{F} and exchange phases \textit{R} used to derive all EBMs. Appendix C gives the specific braiding calculation processes for obtaining the $\sigma^{(3)}_2/\sigma^{(6)}_3$ EBMs. Appendix D introduces GA-enhanced SKA method for topological quantum compilation.
	
	\section{The construction of one-qubit gate}
	
\begin{figure}[h]
		\centering
		\includegraphics[width=0.5\textwidth]{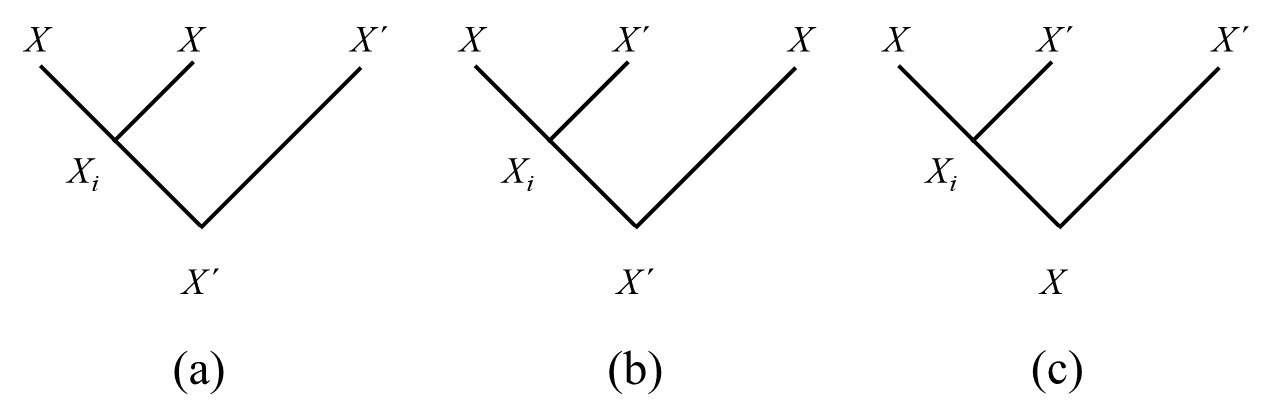}
		\caption{(a) The model $V^{113}_3$. The initial state is composed of two anyons \textit{X} and one $X^{\prime}$ anyon, which are finally fused into one $X^{\prime}$ anyon. (b) The model $V^{131}_3$. Three initial metaplectic anyons \textit{X}, $X^{\prime}$, \textit{X} are fused into one anyon $X^{\prime}$. (c) The model $V^{133}_1$. Three initial metaplectic anyons \textit{X}, $X^{\prime}$, $X^{\prime}$ are fused into one anyon \textit{X}.}
		\label{fig.1}
	\end{figure}

	In topological quantum computing, intermediate fusion states are used to represent qubits. Three anyons can build a one-qubit. Fig. \ref{fig.1} shows the fusion processes of $\{V^{113}_3,V^{131}_3,V^{133}_1\}$ models for constructing one-qubit gates. Specifically, in the initial state, three anyons $(X, X, X^\prime)$ are arranged for $V^{113}_3$, $(X, X^\prime, X)$ for $V^{131}_3$ and $(X, X^\prime, X^\prime)$ for $V^{133}_1$, $X_i$ would be equal to either 1/\textit{Y} (Fig. \ref{fig.1}(a)) or \textit{Y}/\textit{Z} (Fig. \ref{fig.1}(b) and Fig. \ref{fig.1}(c)) according to the fusion rules Eq .\ref{eq.2}. Consequently, for $V^{113}_3$,  binary states through endowing with $X_i$ =1 and \textit{Y} encoding $\ket{0} $ and $\ket{1}$ can be obtained, respectively, and then they are finally fusion into $X^\prime$. These two fusion processes are simply denoted as $\ket{0}=((X,X)_1,X^\prime)_{X^\prime}$, $\ket{1}=((X,X)_Y,X^\prime)_{X^\prime}$.Similarly, the binary states are $\ket{0}=((X,X^\prime)_Y,X)_{X^\prime}$, $\ket{1}=((X,X^\prime)_Z,X)_{X^\prime}$ for $V^{131}_3$ and $\ket{0}=((X,X^\prime)_Y,X^\prime)_X$, $\ket{1}=((X,X^\prime)_Z,X^\prime)_X$ for $V^{133}_1$.
	
	Although both $V^{113}_3$ and $V^{131}_3$ are initially made up of two \textit{X} and one $X^\prime$, and fusing into $X^\prime$ eventually, the different order of fusion results in different intermediate states.  In contrast, $V^{131}_3$ and $V^{133}_1$ share the same intermediate state, yet their initial and final fused anyons differ. These two differences can lead to different EBMs. The EBMs $\sigma^{(3)}_i(i=1,2)$ corresponds to the braiding of the \textit{i} anyon with the \textit{i}+1 anyon, noting that the superscript (3) represents the 3 anyons of the one-qubit, similarly the 6 anyons in the two-qubit EBMs $\sigma^{(6)}_i$. It is easy for $\sigma^{(3)}_1$ to solve, because we only need to braid the first and second anyons (starting from the most left one), and only need to use the \textit{R}-symbols. However, it is slightly complicated to solve $\sigma^{(3)}_2$, which involves braiding the second and third anyons, due to basis transformation which is necessary to use \textit{F}-matrices and \textit{R}-symbols simultaneously. All \textit{F}-matrices and \textit{R}-symbols used in solving EBMs are summarized in Appendix B. The specific solving processes are given in Appendix C. 
	
	For the $\{V^{113}_3,V^{131}_3,V^{133}_1\}$ models, we get the EBMs as follows:
		
			$V^{113}_3:$
		\begin{center}
			$\sigma_1^{(3)} =
			\begin{bmatrix}
				e^{i\frac{3\pi}{4}} & 0\\
				0&e^{i\frac{\pi}{12}}
			\end{bmatrix},$
			$\sigma_2^{(3)} =\frac{1}{3} 
			\begin{bmatrix}
				2e^{i\frac{7\pi}{12}}+e^{i\frac{\pi}{4}} & -\sqrt[]{2}e^{i\frac{7\pi}{12}}+\sqrt[]{2}e^{i\frac{\pi}{4}}\\
				-\sqrt[]{2}e^{i\frac{7\pi}{12}}+\sqrt[]{2}e^{i\frac{\pi}{4}}&e^{i\frac{7\pi}{12}}+2e^{i\frac{\pi}{4}}
			\end{bmatrix}.
			$
		\end{center}
			
			$V^{131}_3:$
			
		\begin{center}
			$\sigma_1^{(3)} =
			\begin{bmatrix}
				e^{i\frac{7\pi}{12}} & 0\\
				0&e^{i\frac{\pi}{4}}
			\end{bmatrix},$
			
			$\sigma_2^{(3)} =\frac{1}{3} 
			\begin{bmatrix}
				e^{i\frac{7\pi}{12}}+2e^{i\frac{\pi}{4}} & -\sqrt[]{2}e^{i\frac{7\pi}{12}}+\sqrt[]{2}e^{i\frac{\pi}{4}}\\
				-\sqrt[]{2}e^{i\frac{7\pi}{12}}+\sqrt[]{2}e^{i\frac{\pi}{4}}&2e^{i\frac{7\pi}{12}}+e^{i\frac{\pi}{4}}
			\end{bmatrix}.
			$
		\end{center}
			
			$V^{133}_1:$
			
		\begin{center}
			$\sigma_1^{(3)} =
			\begin{bmatrix}
				e^{i\frac{7\pi}{12}} & 0\\
				0&e^{i\frac{\pi}{4}}
			\end{bmatrix},$
			
			$\sigma_2^{(3)} =\frac{1}{3} 
			\begin{bmatrix}
				2e^{-i\frac{\pi}{4}}+e^{-i\frac{11\pi}{12}} & -\sqrt[]{2}e^{-i\frac{\pi}{4}}+\sqrt[]{2}e^{-i\frac{11\pi}{12}}\\
				-\sqrt[]{2}e^{-i\frac{\pi}{4}}+\sqrt[]{2}e^{-i\frac{11\pi}{12}}&e^{-i\frac{\pi}{4}}+2e^{-i\frac{11\pi}{12}}
			\end{bmatrix}.
			$
		\end{center}
	
	Standard $H/T$-gate cannot be constructed directly using EBMs of $\{V^{113}_3,V^{131}_3,V^{133}_1\}$ models, similar to the Fibonacci anyon model, only a braidword from the braiding of EBMs approximation to $H/T$-gate can be obtained, which becomes a quantum compilation problem.
	To measure the distance between a braidword and a standard gate, we use the global phase invariant distance \cite{RN35}:
	\begin{equation}
		\small
		d(U_0,U)=\sqrt{1-\frac{|\tr(U_0U^\dagger)|}{2}}
		\label{eq.3},
	\end{equation}
	where $U_0$ represents a matrix of the standard one-qubit gate and $U $ represents the matrix of braidword obtained by the arrangement of EBMs, and $Tr $ denotes the trace of the matrix $U_0U^\dagger$. A smaller distance $d(U_0,U)$ indicates a smaller error between the braidword and the standard gate.
	
	For the one-qubit compilation problem, our solution is the GA-enhanced SKA \cite{RN38}. The SKA promises an exponentially reduced distance at the cost of each five-times increase in length (the number of EBMs) by recursive calls \cite{RN13}. The limitation of the traditional SKA is that 0-order approximations need to be obtained using exhaustive search, which quickly becomes infeasible for basic lengths greater than 14. E. G. Johansen replaced exhaustive search with the Monte Carlo method in the 0-order approximation \cite{RN14}, which resulted in significant performance improvements, including unlimited basic length and a great reduction in time cost. Inspired by Monte Carlo-enhanced SKA, using GA to replace exhaustive search in 0-level approximation, which has similar advantages. For convenience, GA-enhanced SKA is briefly introduced in Appendix D. Table \ref{tab1} shows the 0-level approximate braidwords for $H$/$T$-gate obtained by GA for the $\{\rm{Fibonacci},V^{113}_3,V^{131}_3,V^{133}_1\}$. We run the GA three times and select the individual with the highest fitness for calculations. By recursively calling SKA, the 1-, 2-, and 3-level approximations for \textit{H}- and \textit{T}-gates can be further obtained.
	
	\begin{figure}[h]
		\centering
		\includegraphics[width=0.5\textwidth]{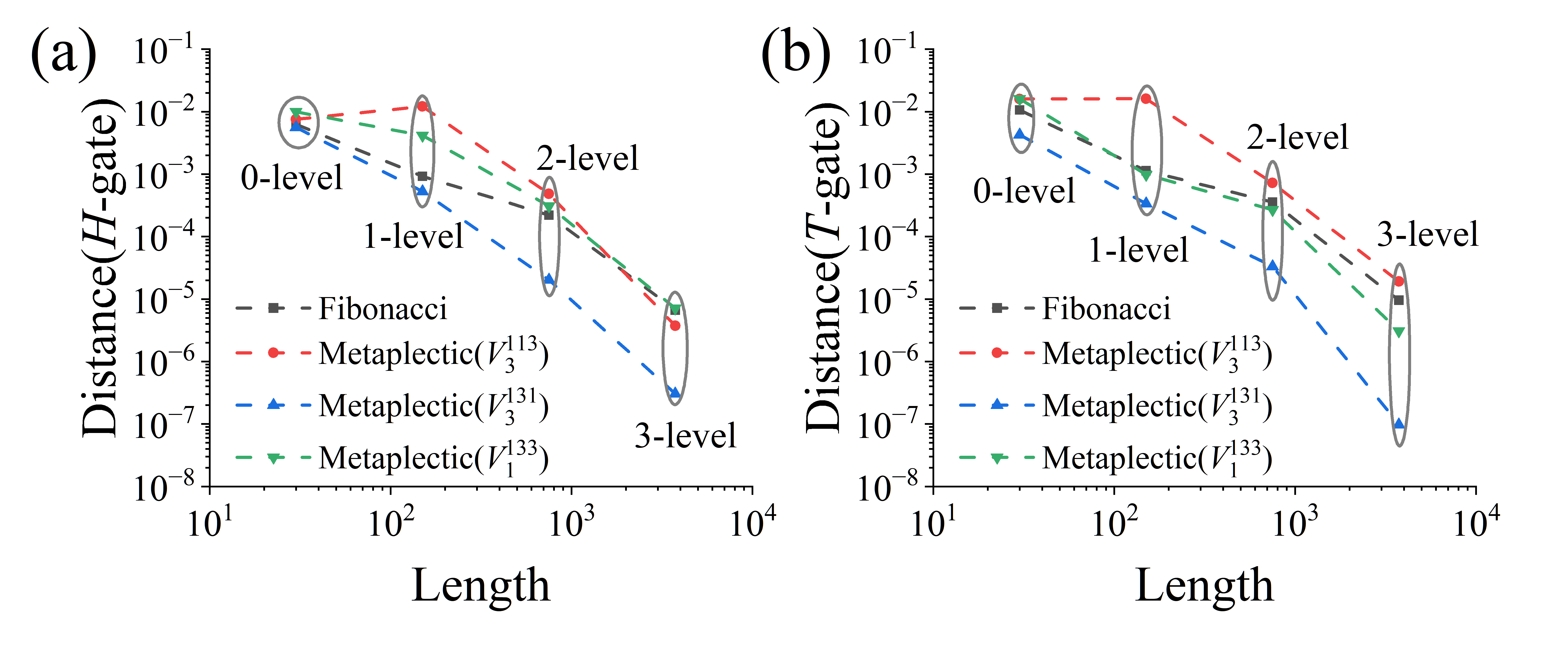}
		\caption{The $\{V^{113}_3,V^{131}_3,V^{133}_1\} $ model uses GA-enhanced SKA to obtain 0-, 1-, 2-, and 3-level approximations of the (a) $H$-gate and (b) $T$-gate. The basic length $L_0 $=30 is set.}
		\label{fig.2}
	\end{figure}
	
	Setting the same parameters in GA, the results shown in Fig. \ref{fig.2} by running the GA-enhanced SKA algorithm on $\{\rm{Fibonacci},V^{113}_3,V^{131}_3,V^{133}_1\}$ four models can be obtained, respectively. For the approximation of the standard \textit{H}-gate, Fig. \ref{fig.2}(a) shows that the $V^{131}_3$ model performs best at all levels of approximation, and $V^{113}_3/V^{133}_1$  can achieve a comparable distance ($ 10^{-5}$) with the Fibonacci model at the 3-level approximation. The threshold theorem requires that the error between the braidword and the standard qubit gate is less than 1\% \cite{RN39,RN40,RN41,RN42,RN43,RN44}, which means that our $d(U_0,U)$ from the standard gate needs to be less than  ($ 10^{-2}$). For the $V^{131}_3$ model, we only need to achieve the 1-level approximation (corresponding to $30\times5^1$ braid operations) to meet the computational requirements. For the $\{\rm Fibonacci,V^{113}_3,V^{133}_1\}$ model, we need to achieve a 2-level approximation (corresponding to $30\times5^2$ braid operations), so choosing $V^{131}_3$ for calculations can reduce a number of unnecessary braid operations. Fig. \ref{fig.2}(b) demonstrates the best performance when approximating the standard \textit{T}-gate for the model $V^{131}_3$. Similar to the \textit{H}-gate approximation, to meet the requirements of the threshold theorem, we need to achieve a 1-order approximation for $V^{131}_3$ and a 2-order approximation for $\{\rm{Fibonacci},V^{113}_3,V^{133}_1\}$.
	
	\begin{widetext}
		
		\begin{table}[h]
			\centering
			\caption{ The braidwords of four models with the global phase invariant distance of \textit{H}/\textit{T}-gate at basic length 30. The A/B/C/D represents $\sigma_1^{(3)}/\sigma_2^{(3)}/\sigma_1^{(3)-1}/\sigma_2^{(3)-1}$, respectively.}
			\begin{tabular}{ccc}
				\hline
				\hline
				Models&braidwords&distance\\
				\hline
				$\rm{Fibonacci(\textit{H}-gate)}$&CDADDADCBADDADDDDCDADADADADADD&0.006268\\
				$V^{113}_3(H-gate)$&DDDDADADDDCDCDCBADDDCDDDCDADAB&0.007563\\
				$V^{131}_3(H-gate)$&ADDADABBABBBADCDDABCDABCCCDDDA&0.005593\\
				$V^{133}_1(H-gate)$&DDCDADDDDCDADADCDDCDCDDADDADDC&0.009955\\
				$\rm{Fibonacci(\textit{T}-gate)}$&ADDDCDDADDADADCDCDADDADDDDDCCD&0.010634\\
				$V^{113}_3(T-gate)$&DDDCDCDAADDDADDDCDCDCDCDDDADDC&0.015955\\
				$V^{131}_3(T-gate)$&DCBABADCCDAABCCCBBBADABCDAAABC&0.004259\\
				$V^{133}_1(T-gate)$&BCDCBCDDDDADADCCCDCDDCDDDDDDDA&0.015954\\
				\hline
				\hline
			\end{tabular}
			\label{tab1}
		\end{table}
	\end{widetext}
	
	\section{The construction of two-qubit gates}
	
\begin{figure}[h]
		\centering
		\includegraphics[width=0.45\textwidth]{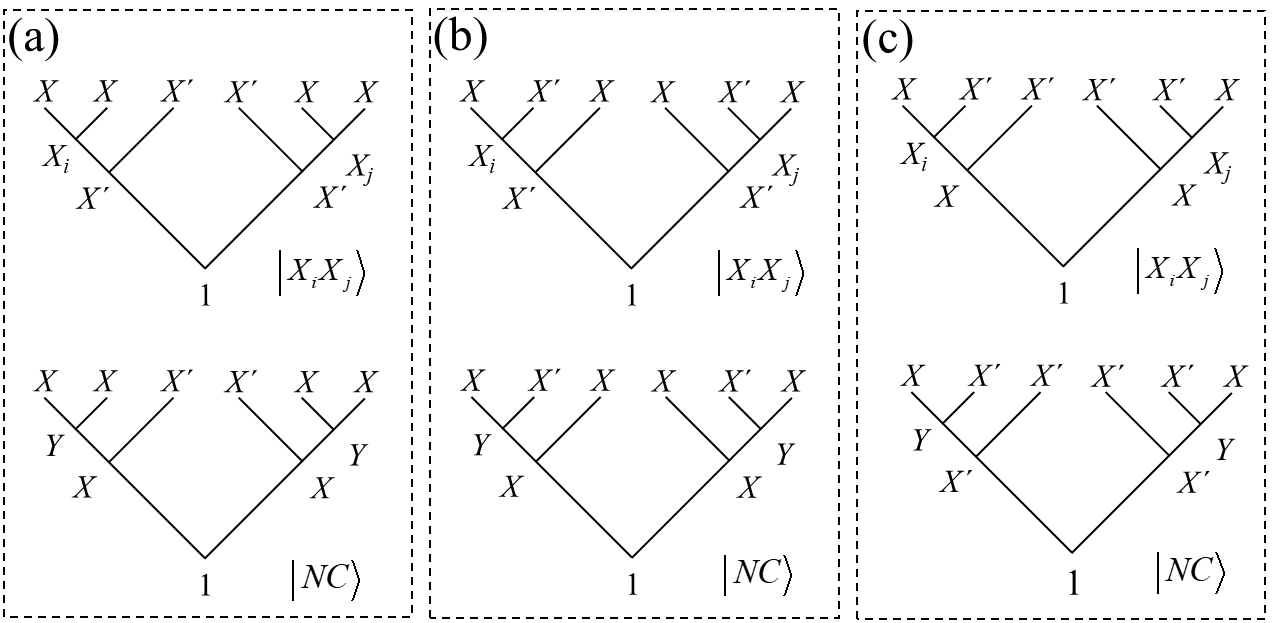}
		\caption{The top/bottom figures correspond to the computational/non-computational states. (a) The two-qubit model of $V^{113}_3$; (b) The two-qubit model of $V^{131}_3$;(c) The two-qubit model of $V^{133}_1$.}
		\label{fig.3}
	\end{figure}

	Encoding two-qubit requires the use of 6 anyons for our model. Fig. \ref{fig.3}(a) shows the computational and non-computational states of the two-qubit model for $V^{113}_3$. We first arrange four anyons \textit{X} and two anyons $X^\prime$, then finally fuse them into the vacuum. According to the fusion rules, four computational states $\ket{X_iX_j}(\ket{11},\ket{1Y},\ket{Y1},\ket{YY})$ will be generated in the fusion processes, and we can encode them corresponding as $(\ket{00},\ket{01},\ket{10},\ket{11})$, a non-computational state  $\ket{NC}$ will produce simultneously,  similar to the Fibonacci model with 6 anyons code. Similarly, Fig. \ref{fig.3}(b) shows the computational state $(\ket{YY},\ket{YZ},\ket{ZY},\ket{ZZ})$ , which also correspond to $(\ket{00},\ket{01},\ket{10},\ket{11})$ and non-computational state for $V^{131}_3$, and Fig. \ref{fig.3}(c) shows the computational state $(\ket{YY},\ket{YZ},\ket{ZY}$ and non-computational states for $V^{133}_1$. Note that   $V^{131}_3$ and $V^{133}_1$ have the same encoding intermediate state, but differences in the initial anyons result in different EBMs. 
	
	For encoding two-qubit, in the case of basic vectors $(\ket{NC},\ket{11},\ket{1Y},\ket{Y1},\ket{YY})$ for $V^{113}_3$,
	$(\ket{NC},\ket{YY},\ket{YZ},\ket{ZY},\ket{ZZ})$ for $V^{131}_3(V^{133}_1)$, we obtain the EBMs as follows:
	
	\begin{widetext}
		$V^{113}_3:$ 
		\begin{center}
			$\sigma_1^{(6)}=R^{11}_2\oplus(\sigma_1^{(3)}\otimes I_2),
			\sigma_2^{(6)} =R^{13}_2\oplus(\sigma_2^{(3)}\otimes I_2),$	
			$\sigma_3^{(6)} =\frac{1}{2}
			\begin{bmatrix}
				e^{-i\frac{\pi}{4}}+e^{-i\frac{11\pi}{12}}&0&0&0&-e^{-i\frac{\pi}{4}}+e^{-i\frac{11\pi}{12}}\\
				0&2e^{-i\frac{\pi}{4}}&0&0&0\\
				0&0&2e^{-i\frac{11\pi}{12}}&0&0\\
				0&0&0&2e^{-i\frac{11\pi}{12}}&0\\
				-e^{-i\frac{\pi}{4}}+e^{-i\frac{11\pi}{12}}&0&0&0&e^{-i\frac{\pi}{4}}+e^{-i\frac{11\pi}{12}}
			\end{bmatrix},$
			$\sigma_4^{(6)} =R^{31}_2\oplus(I_2\otimes \sigma_2^{(3)}),
			\sigma_5^{(6)} =R^{11}_2\oplus(I_2\otimes  \sigma_1^{(3)}).$\\
		\end{center}
		$V^{131}_3:$
		\begin{center}
			$\sigma_1^{(6)}=R^{13}_2\oplus(\sigma_1^{(3)}\otimes I_2),
			\sigma_2^{(6)}=R^{31}_2\oplus(\sigma_2^{(3)}\otimes I_2),$\\
			$\sigma_3^{(6)}=\frac{1}{2}
			\begin{bmatrix}
				e^{i\frac{3\pi}{4}}+e^{i\frac{\pi}{12}}&-e^{i\frac{3\pi}{4}}+e^{i\frac{\pi}{12}}&0&0&0\\
				-e^{i\frac{3\pi}{4}}+e^{i\frac{\pi}{12}}&e^{i\frac{3\pi}{4}}+e^{i\frac{\pi}{12}}&0&0&0\\
				0&0&2e^{-i\frac{\pi}{12}}&0&0\\
				0&0&0&2e^{-i\frac{\pi}{12}}&0\\
				0&0&0&0&2e^{i\frac{3\pi}{4}}
			\end{bmatrix},$\\
			$
			\sigma_4^{(6)}=R^{13}_2\oplus(I_2\otimes \sigma_2^{(3)}),
			\sigma_5^{(6)}=R^{31}_2\oplus(I_2\otimes  \sigma_1^{(3)}).$
		\end{center}
		$V^{133}_1:$
		\begin{center}
			$
			\sigma_1^{(6)}=R^{13}_2\oplus(\sigma_1^{(3)}\otimes I_2),
			\sigma_2^{(6)}=R^{33}_2\oplus(\sigma_2^{(3)}\otimes I_2),$
			\\
			$
			\sigma_3^{(6)}=\frac{1}{2}
			\begin{bmatrix}
				e^{-i\frac{\pi}{4}}+e^{-i\frac{11\pi}{12}}&-e^{-i\frac{\pi}{4}}+e^{-11i\frac{\pi}{12}}&0&0&0\\
				-e^{-i\frac{\pi}{4}}+e^{-11i\frac{\pi}{12}}&e^{-i\frac{\pi}{4}}+e^{-i\frac{11\pi}{12}}&0&0&0\\
				0&0&2e^{-i\frac{11\pi}{12}}&0&0\\
				0&0&0&2e^{-i\frac{11\pi}{12}}&0\\
				0&0&0&0&2e^{-i\frac{\pi}{4}}\\
			\end{bmatrix},$
			\\
			$
			\sigma_4^{(6)}=R^{33}_2\oplus(I_2\otimes \sigma_2^{(3)}),
			\sigma_5^{(6)}=R^{31}_2\oplus(I_2\otimes  \sigma_1^{(3)}).
			$
		\end{center}	
	\end{widetext}
	
	For the $\sigma_1^{(6)}/\sigma_5^{(6)}$ is easy to get, we only need to braid the first/last two anyons, only \textit{R}-symbols need to be used. The solution of the $\sigma_2^{(6)}/\sigma_3^{(6)}/\sigma_4^{(6)}$ is much more complicated, and it needs to change the basis to a braid corresponding to two anyons, \textit{F}-matrices and \textit{R}-symbols should be used simultaneously, and the specific solution process for $\sigma_3^{(6)}$ is referred to Appendix C.
	
	For the ${V^{113}_3,V^{131}_3,V^{133}_1}$ models, there is a direct product relationship between the EBMs $\sigma_1^{(6)}/\sigma_2^{(6)}/\sigma_4^{(6)}/\sigma_5^{(6)} $ of the two-qubit gates and the EBMs $\sigma_1^{(3)}/\sigma_2^{(3)}$ of the one-qubit, but there is no such relationship for $\sigma_3^{(6)}$, which indicates the entanglement of two qubits.
	
	Among the two-qubit gates, CNOT-gate is the key, and it can be combined with \textit{H}-gate and \textit{T}-gate to realize universal quantum computing. The braidword formed using the five EBMs of $\{V^{113}_3,V^{131}_3,V^{133}_1\}$ is a 5-dimensional matrix, and we write the braidword as $B=M_{11}\bigoplus A$ so that $M_{11}$ corresponds to the non-computational space and $|M_{11}|=\sqrt{M_{11}M_{11}^\dagger} \approx1$ \cite{RN19}, \textit{A} corresponds to the 4-dimensional computational space matrix. To construct the CNOT-gate, we try to approximate a local equivalence class of the CNOT-gate with the \textit{A}-matrix \cite{RN45}. Following the work \cite{RN46}, we convert the \textit{U} (which can be a standard two-qubit gate or a computational matrix \textit{A} in a braidword) from a standard computational basis to a Bell basis,
	
	\begin{equation}
		\begin{split}	
			U_B=Q^\dagger UQ,Q=\frac{1}{\sqrt{2}}
			\begin{bmatrix}
				1&0&0&i\\
				0&i &1&0\\
				0&i &-1&0\\
				1&0&0&-i
			\end{bmatrix}
		\end{split}
		\label{eq.4},
	\end{equation}
	
	then obtain three real parameters $g_1, g_2, g_3$ called local invariants:
	
	\begin{equation}
		\begin{split}
			g_1 &=Re{\{\frac{tr^2(m_u)}{16\cdot U}}\}, \\
			g_2 &=Im{\{\frac{tr^2(m_u)}{16\cdot U}}\}, \\
			g_3 &={\frac{tr^2(m_u)-tr(m^2_u)}{4\cdot U}},
		\end{split}
		\label{eq.5}
	\end{equation}
	
	where $m_u=U_B^TU_B$.These formulas provided in the reference \cite{RN20} are used to measure the distance between the braidword and the local invariant of the CNOT-gate,
	
	\begin{equation}
		d^{CNOT}(A)=\sum_{i=1}^{3}\triangle g^2_i,
		\triangle g_i=|g_i(A)-g_i(CNOT)|,
	\end{equation}
	
	where \textit{A} is the computational matrix in the braidword, and the $g_1, g_2, g_3$ of CNOT-gate can be obtained directly by Eq .\ref{eq.4} and Eq .\ref{eq.5}, they are
	
	\begin{center}
		$g_1(CNOT)=0,g_2(CNOT)=0,g_3(CNOT)=1.$
	\end{center}
	
	When the three local invariants of the \textit{A}-matrix are close to the values of the CNOT-gate, this indicates that the braidword \textit{B} are functionally equivalent with the CNOT-gate. The two operations can be interconverted through single-qubit operations (local operations). 
	
	At the same time, to ensure that the \textit{A}-matrix is approximately unitary, we also make unitary measurements of \textit{A}:
	
	\begin{equation}
		d^U=Tr(\sqrt{a^\dagger a}),\quad a=A^\dagger A-I,
	\end{equation}
	where \textit{I} is a 4-dimensional identity matrix.
	
	\begin{figure}[h]
		\centering
		\includegraphics[width=0.47\textwidth]{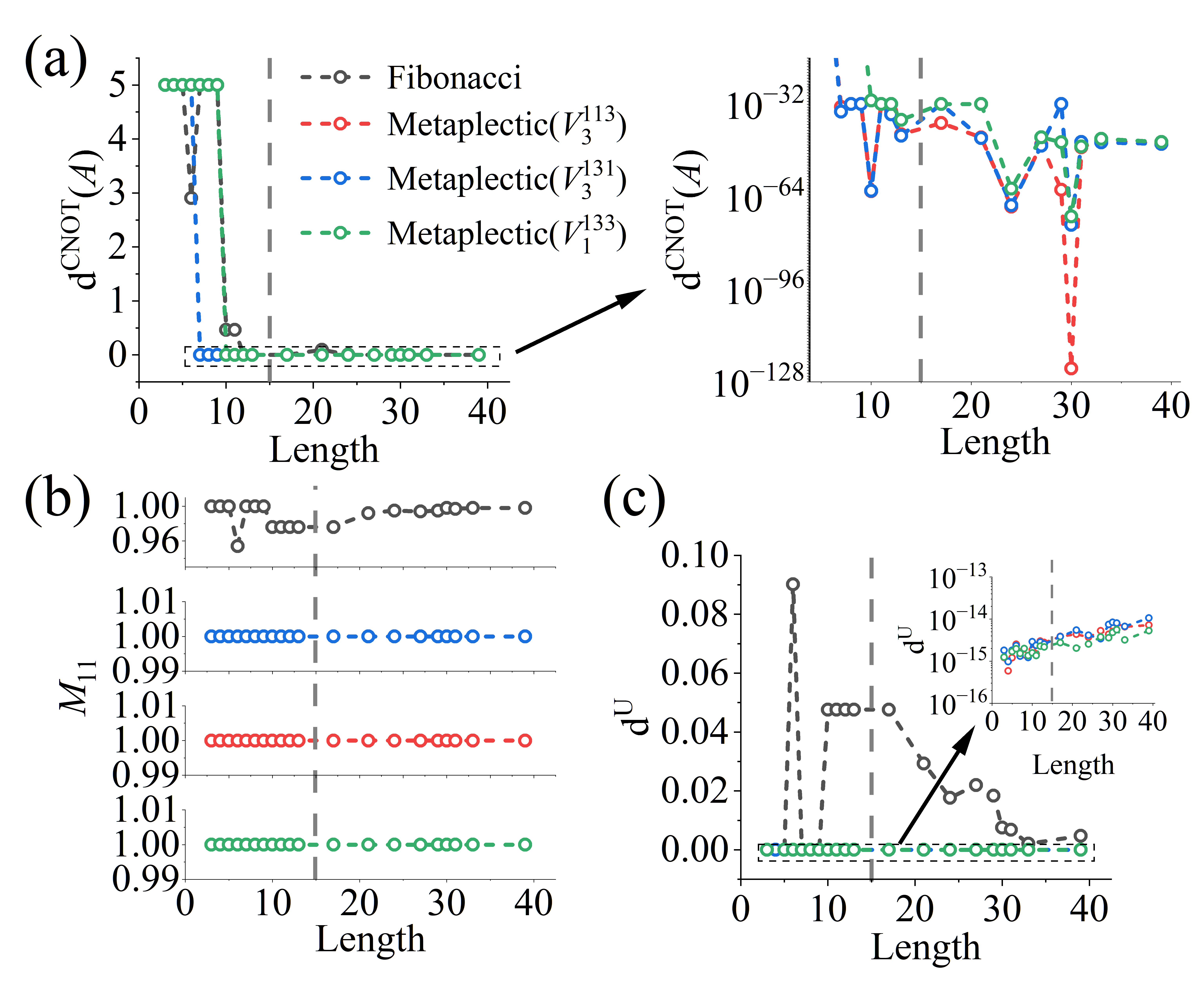}
		\caption{The distance of local equivalence class [CNOT] and unitary measurement without inverse EBMs. (a) The distance between the $\{\rm Fibonacci,V^{113}_3,V^{131}_3,V^{133}_1\}$ models and the local equivalence class [CNOT] at different lengths. We use exhaustive search at short lengths, and GA at long lengths. The gray dashed line distinguishes the two methods. (b)$M_{11}$ for $\{\rm Fibonacci,V^{113}_3,V^{131}_3,V^{133}_1\}$. (c) Unitary measurement of \textit{A}-matrix for $\{\rm Fibonacci,V^{113}_3,V^{131}_3,V^{133}_1\}$.}
		\label{fig.4}
	\end{figure}
	
\begin{figure}[h]
		\centering
		\includegraphics[width=0.47\textwidth]{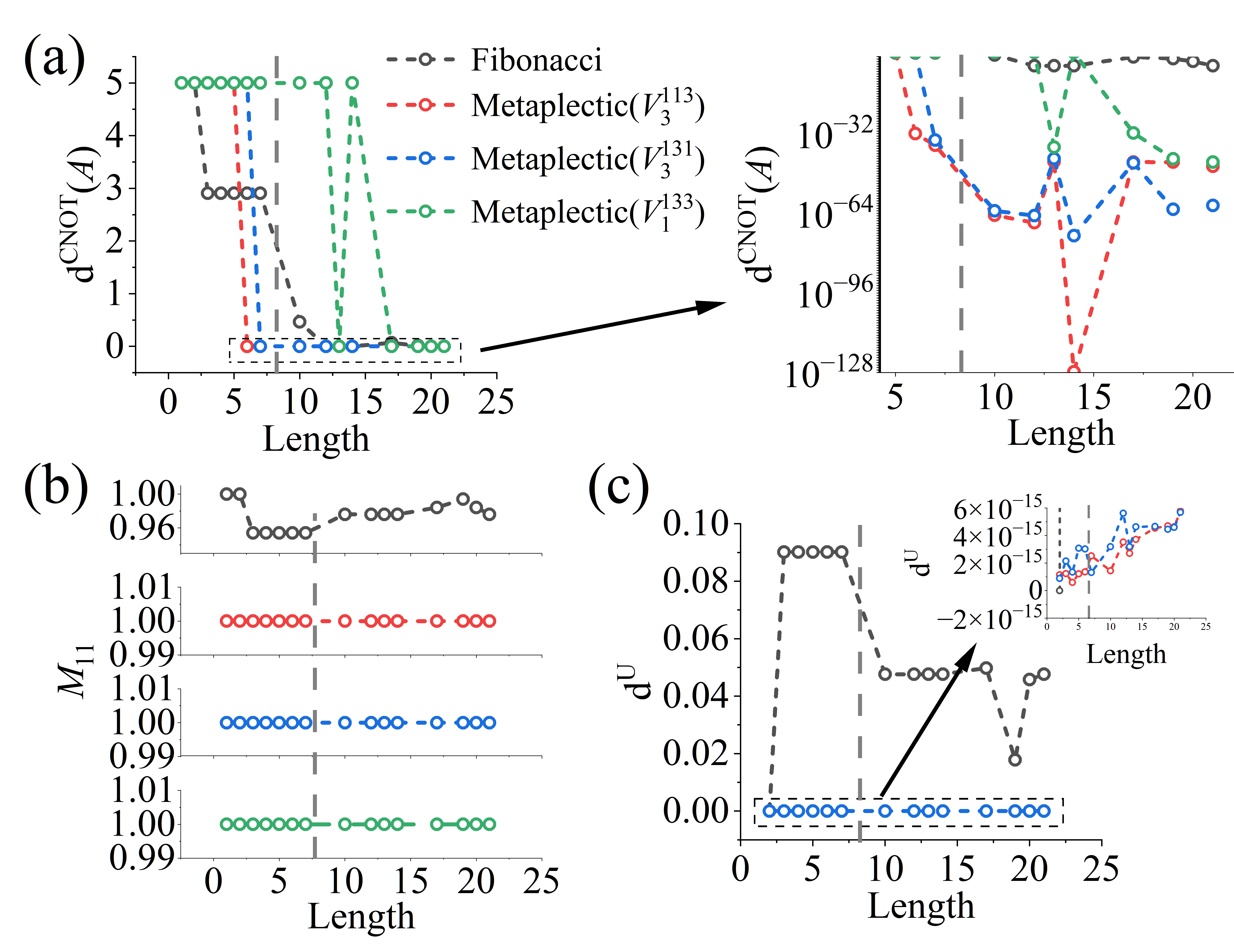}
		\caption{The distance of local equivalence class [CNOT] and unitary measurement with inverse EBMs. (a) The distance between the $\{\rm Fibonacci,V^{113}_3,V^{131}_3,V^{133}_1\}$ models and the local equivalence class [CNOT] at different lengths. We use exhaustive search at short lengths, and GA at long lengths. The gray dashed line distinguishes the two methods. (b)$M_{11}$ for $\{\rm Fibonacci,V^{113}_3,V^{131}_3,V^{133}_1\}$. (c) Unitary measurement of \textit{A}-matrix for $\{\rm Fibonacci,V^{113}_3,V^{131}_3,V^{133}_1\}$.}
		\label{fig.5}
	\end{figure}

	Note the data of the Fibonacci model in Fig. \ref{fig.4} and Fig. \ref{fig.5} comes from \cite{RN20}. For exhaustive search based on 5/10 kinds of EBMs, the length reaches 13/7. The inverse matrix is not added to our calculations so that the exhaustive search can reach a longer length. The GA is used for the length larger than the upper limit of the exhaustive search, and the results are shown in Fig. \ref{fig.4}. The results show that the $V^{113}_3$ approximation of a local equivalence class for CNOT-gate is satisfactory, even reaches close to $ 10^{-128}$ when the length is 30, which is far less than the other three models (Fig. \ref{fig.4}(a)). The distance of the local equivalence class decreases rapidly (reaching below $ 10^{-32}$) when the length is greater than 10 for $\{V^{113}_3,V^{131}_3,V^{133}_1\}$, far less than the Fibonacci model. The value of $M{11}$ can remain in a straight line for $\{V^{113}_3,V^{131}_3,V^{133}_1\}$, but the Fibonacci model produces some considerable fluctuations. The unitary measurement of $\{V^{113}_3,V^{131}_3,V^{133}_1\}$ for the \textit{A}-matrix can always be kept at an ideal value (below $10^{-14} $), which is generally better than the Fibonacci model.
	The calculation results of adding the inverse matrices are shown in Fig. \ref{fig.5}. The conclusion is consistent with no inverse matrices, $V^{113}_3$ giving the best result in all models. Compared with the case without the inverse matrices, $V^{113}_3$ only needs \textit{l} = 14 to reach $10^{-128}$ approximately.  At the same time, we find that when the inverse matrices are added, three models   find a braidword with a distance equal to 0 (below $10^{-128}$) from the local equivalence class [CNOT] when the length is 20. We show these three braidwords in Table \ref{tab2}. The $M_{11}$ of  $\{V^{113}_3,V^{131}_3,V^{133}_1\}$ always remains equal to 1, which $M_{11}$ of the Fibonacci model is often far from 1 for comparison. For unitary measurements of the \textit{A}-matrix, the results indicate that three models $\{V^{113}_3,V^{131}_3,V^{133}_1\}$ can reach an excellent level (below $6\times10^{-15}$), superior to the Fibonacci model.
	
	\begin{widetext}
		
		\begin{table}[h]
			\centering
			\caption{ The braidwords of three models with the local equivalent class [CNOT] distance equal 0 at length 20, and corresponding M11 and unitary measurements of A-matrix. The A/B/C/D/E/F/G/H/I/J represents $\sigma_1^{(6)}/\sigma_2^{(6)}/\sigma_3^{(6)}/\sigma_4^{(6)}/\sigma_5^{(6)}/\sigma_1^{(6)-1}/\sigma_2^{(6)-1}/\sigma_3^{(6)-1}/\sigma_4^{(6)-1}/\sigma_5^{(6)-1}$, respectively.}
			\begin{tabular}{ccccc}
				\hline
				\hline
				Model&braidword&distance&$M_{11}$&Unitary measurement(\textit{A})\\
				\hline
				$V^{113}_3$&BBIFBDAAHFJBAHBHBBJA&0&1&$4.629\times10^{-15}$\\
				$V^{131}_3$&GFEAGJCBAAHHBCBBBJBJ&0&1&$4.586\times10^{-15}$\\
				$V^{133}_1$&DGIGJHBFBEFFCBFBHBFE&0&1&$2.554\times10^{-15}$\\
				\hline
				\hline
			\end{tabular}
			\label{tab2}
		\end{table}
	\end{widetext}
	
	\section{Differences between conventional and unconventional encoding braiding processes}
	
	For the braiding process of anyons to have topological protection, the braiding operation $\sigma_i$ needs to satisfy the relations of the Artin braid group\cite{RN49}:
	\begin{equation}
		\centering
		\begin{aligned}
			&\sigma_i\sigma_j=\sigma_j\sigma_i\quad for\quad\abs{i-j}\geq2,\\
			&\sigma_i\sigma_{i+1}\sigma_i=\sigma_{i+1}\sigma_i\sigma_{i+1},
		\end{aligned}
		\label{eq.8}
	\end{equation}
	
	where $\sigma_i$ represents the exchange of the i anyon and i+1 anyon. The \textit{H}-/\textit{T}-gate, and CNOT-gate can be constructed with excellent performance using the EBMs of the $\{V^{113}_3,V^{131}_3,V^{133}_1\}$. Unfortunately, none of their EBMs satisfy the Artin braid group relations Eq. \ref{eq.8}. This implies that the braiding processes of these three models are not topologically protected. However, this does not imply that our derivation of EBMs is incorrect. The EBMs obtained for the $V^{111}_1$ model using conventional encoding comply with the Artin relations Eq. \ref{eq.8}. We will now demonstrate this analytical process.
	
	\begin{figure}[h]
		\centering
		\includegraphics[width=0.45\textwidth]{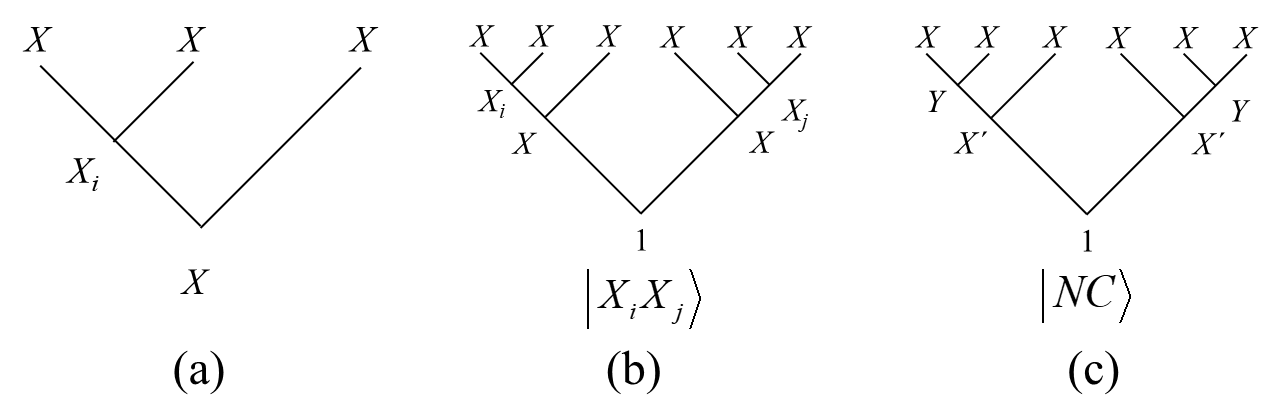}
		\caption{(a) The one-qubit model of $V^{111}_1$. The initial state is composed of three \textit{X} anyons, which are fused into one \textit{X} anyon finally. (b) The two-qubit model  of $V^{111}_1$. Six initial \textit{X} anyons are fused into vacuum eventually. (c) The non-computational state of two-qubit $V^{111}_1$ model .}
		\label{fig.6}
	\end{figure}
	
	Based on the fusion process in Fig. \ref{fig.6}, we define the one and two quantum logical qubit under the $V^{111}_1$. Fig. \ref{fig.6}(a) shows the fusion process of a one-qubit. Similar to $\{V^{113}_3,V^{131}_3,V^{133}_1\}$, we can represent one-qubit by $\ket{X_i}(\ket{1},\ket{Y})$ as $(\ket{0},\ket{1}$. Fig. \ref{fig.6}(b) corresponds to the fusion process of two-qubit. For two-qubit, it can be represented by $\ket{X_iX_j}(\ket{11},\ket{1Y},\ket{Y1},\ket{YY})$ corresponding as $(\ket{00},\ket{01},\ket{10},\ket{11})$. Fig. \ref{fig.6}(c) represents the non-computational state. We have precisely solved the EBMs for one-qubits and two-qubits in this model:
	
	\begin{widetext}
		EBMs of one-qubit:
		\begin{center}
			
			$\sigma_1^{(3)} =
			\begin{bmatrix}
				e^{i\frac{3\pi}{4}} & 0\\
				0&e^{i\frac{\pi}{12}}
			\end{bmatrix},
			\sigma_2^{(3)} =\frac{1}{3} 
			\begin{bmatrix}
				e^{i\frac{3\pi}{4}}+2e^{i\frac{\pi}{12}} & -\sqrt[]{2}e^{i\frac{3\pi}{4}}+\sqrt[]{2}^{i\frac{\pi}{12}}\\
				-\sqrt[]{2}e^{i\frac{3\pi}{4}}+\sqrt[]{2}e^{i\frac{\pi}{12}}&2e^{i\frac{3\pi}{4}}+e^{i\frac{\pi}{12}}
			\end{bmatrix},
			$
		\end{center}
		
		EBMs of two-qubit:
		\begin{center}
			\begin{center}
				$\sigma_1^{(6)}=R^{11}_2\oplus(\sigma_1^{(3)}\otimes I_2),
				\sigma_2^{(6)} =R^{11}_2\oplus(\sigma_2^{(3)}\otimes I_2),$
				
				$\sigma_3^{(6)} =\frac{1}{2}
				\begin{bmatrix}
					e^{i\frac{3\pi}{4}}+e^{i\frac{\pi}{12}}&0&0&0&-e^{i\frac{3\pi}{4}}+e^{i\frac{\pi}{12}}\\
					0&2e^{i\frac{3\pi}{4}}&0&0&0\\
					0&0&2e^{i\frac{\pi}{12}}&0&0\\
					0&0&0&2e^{i\frac{\pi}{12}}&0\\
					-e^{i\frac{3\pi}{4}}+e^{i\frac{\pi}{12}}&0&0&0&e^{i\frac{3\pi}{4}}+e^{i\frac{\pi}{12}}
				\end{bmatrix},$
				
				$\sigma_4^{(6)} =R^{11}_2\oplus(I_2\otimes \sigma_2^{(3)}),
				\sigma_5^{(6)} =R^{11}_2\oplus(I_2\otimes  \sigma_1^{(3)}).$\\
			\end{center}
		\end{center}
	\end{widetext}
	
	It is easy to prove that the EBMs of the $V^{111}_1$ satisfy Eq. \ref{eq.8}, which indicates that the braiding process of $V^{111}_1$ has the property of topological protection. However, these EBMs cannot be used to construct the set of gates $\{H,T,CNOT\}$. 
	
	\begin{figure}[h]
		\centering
		\includegraphics[width=0.45\textwidth]{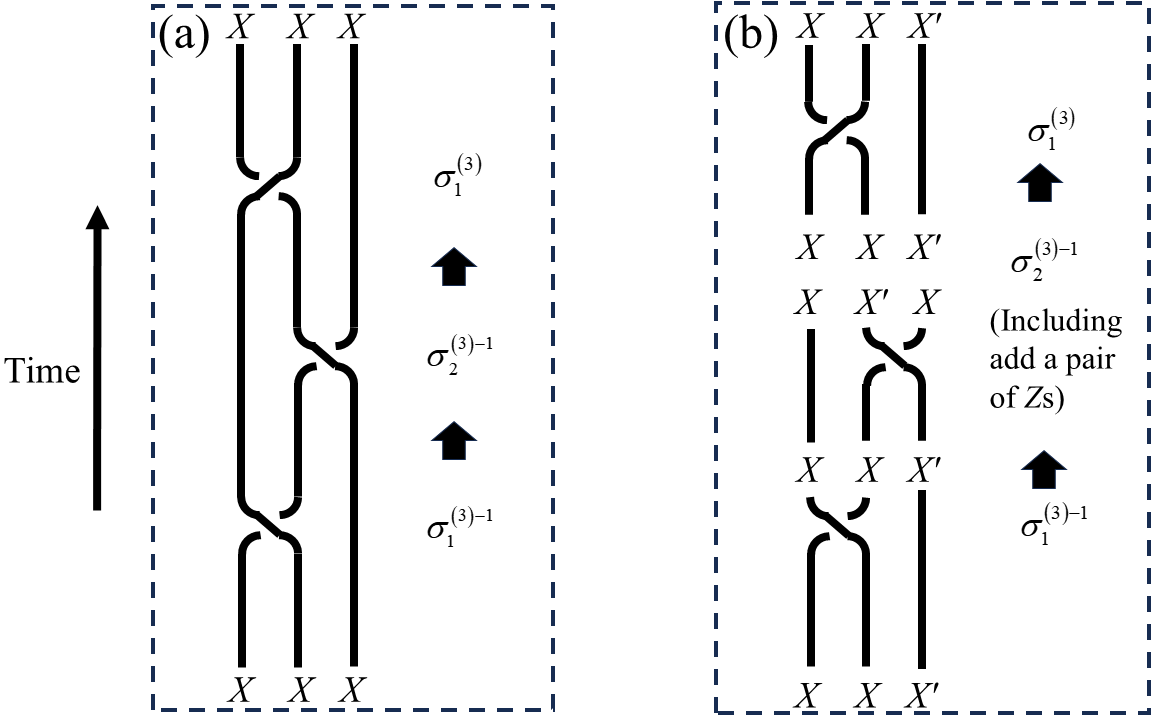}
		\caption{The braiding processes of worldlines for (a) conventional encoding ($V^{111}_1$ model) and (b) unconventional encoding ($V^{113}_3$ model). The braidword is $\sigma^{(3)-1}_1$ $\sigma^{(3)-1}_2$ $\sigma^{(3)}_1$. The time direction is from bottom to top. }
		\label{fig.7}
	\end{figure}
	
    Fig. \ref{fig.7} demonstrates the differences in the braiding processes between the conventional encoding ($V^{111}_1$ model) and unconventional encoding ($V^{113}_3$ model) with the braiding word $\sigma^{(3)-1}_1$ $\sigma^{(3)-1}_2$ $\sigma^{(3)}_1$. Fig. \ref{fig.7}(a) illustrates the braiding process using conventional encoding (based on the $V^{111}_1$ model with three $X$ anyons). Crucially, this conventional encoding process does not require the addition of $Z$ anyon for fusion. Fig. \ref{fig.7}(b) illustrates the braiding process using unconventional encoding (based on the $V^{113}_3$ model with two $X$ anyons and one $X^{\prime}$ anyon). Braiding the two $X$ anyons using $\sigma^{(3)-1}_1$/$\sigma^{(3)}_1$ is straightforward and does not require adding $Z$ anyon for fusion. However, performing $\sigma^{(3)-1}_2$ to braid the $X$ anyon and $X^{\prime}$ anyon causes their positions to swap. To restore the original anyon order, a pair of $Z$ anyons must be added. This operation demonstrably cuts the worldlines and thus violates the Artin relations.
	
	\section{Generalize to N-qubit case}
	
\begin{figure}[t]
		\centering
		\includegraphics[width=0.45\textwidth]{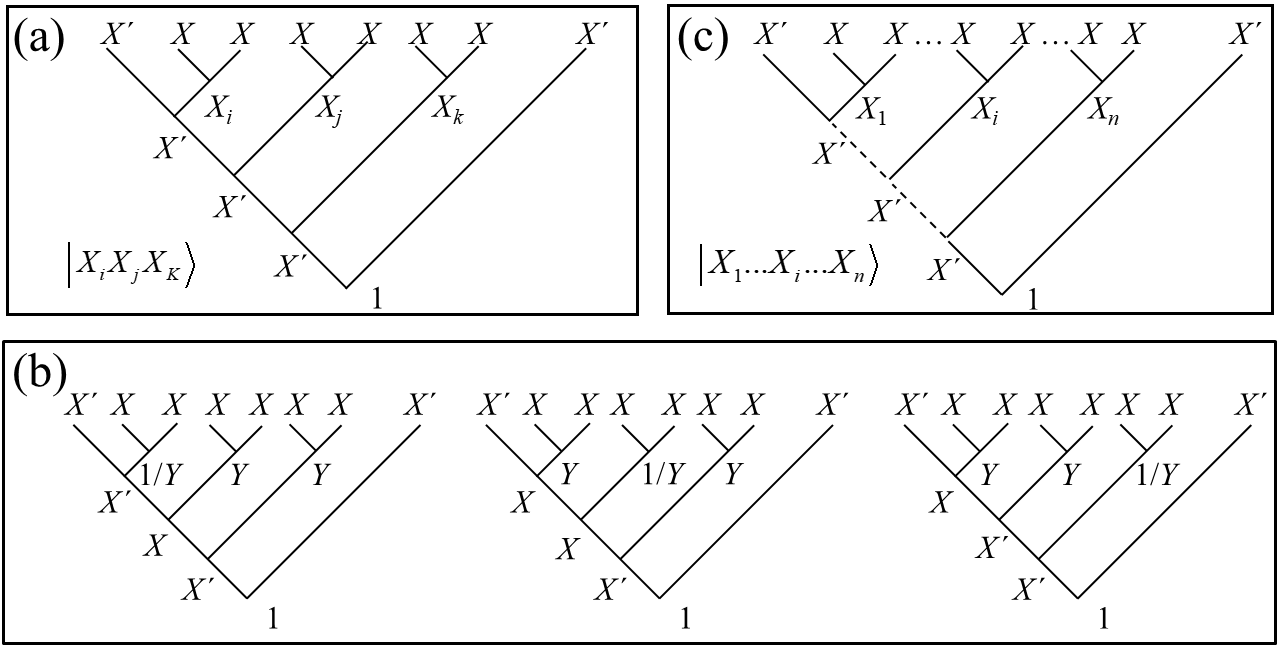}
		\caption{(a) Three-qubit system of the $V^{113}_3$ model. (b) Non-computational states of the $V^{113}_3$ model in the three-qubit system. (c) \textit{N}-qubit system of the $V^{113}_3$ model.}
		\label{fig.8}
	\end{figure}

	We demonstrate that Majorana fermions — corresponding to the Ising anyon model in $SU(2)_2$ topological field theory — can theoretically form an $N$-logical qubit system through multi-body interactions using $N$+1 physical qubits (corresponding to 2($N$+1) Majorana fermions) \cite{RN47}. Consequently, it becomes imperative to explore methods for extending the metaplectic anyon model to \textit{N}-qubit systems. Our model can be extended to \textit{N}-qubits, according to the approach in reference \cite{RN21}. Fig.\ref{fig.8} (a) illustrates a tree diagram for constructing a three-qubit system in the $V^{113}_3$, where the physical qubit $\ket{X_iX_jX_k}$ $(X_i,X_j,X_k\in\{1,Y\})$ correspond to logical qubit $\ket{abc}$ $(a,b,c\in\{0,1\})$, while  Fig.\ref{fig.8} (b) depicts the six non-computational states. In this system, the three pairs of \textit{X} anyons in the middle of the fusion result correspond to three qubits, which fuse into the vacuum eventually. Following our previous methodology, EBMs could theoretically be derived by utilizing \textit{F}-matrices, \textit{R}-symbols, and the insertion of a pair of \textit{Z} anyons at positions $X$ and $X^\prime$. These EBMs would then serve as building blocks for standard quantum gates. However, in the three-qubit case, the EBMs attain a dimensionality of 14 (comprising an 8-dimensional computational subspace and a 6-dimensional non-computational subspace), which poses significant challenges for exact analytical solutions.
	
	We propose the following concrete steps for constructing standard three-qubit quantum gates using 8 metaplectic anyons (based on the $V^{113}_3$ model):
	
	(\romannumeral1) Initial configuration: Position two $X^\prime$ anyons at the endpoints, with six $X$ anyons arranged between them.
	
	(\romannumeral2) State encoding: Identify all fusion paths resulting in the vacuum charge 1 finally. The possible number of fusion channels corresponds to the dimension of the EBMs. Encode the computational basis states as those where the fusion outcome of the first three, first five, and first seven anyons (counting from the initial sorting of anyon) are $X^\prime$, as illustrated in Fig.\ref{fig.8} (a). The remaining fusion paths  correspond to non-computational states, shown in Fig.\ref{fig.8} (b).
	
	(\romannumeral3) Generating braiding operations: Derive all elementary braiding operations $\sigma_i (1\leq i \leq 7)$  using $F$-move and $R$-move. These $\sigma_i$s generate the corresponding set of EBMs.
	
	(\romannumeral4) Standard gate of approximation: Using exhaustive search or GA to find a braidword $B$ (a sequence of EBMs) approximating the target three-qubit gate. Decompose the braidword representation matrix $B$ into a direct sum: $B=M\oplus A$. Here, $B$ has 14-dimension, the non-computational block $A$ has 6-dimension, the computational block $M$ (8-dimension) approximates the standard quantum gate. Crucially, this direct sum decomposition requires the off-diagonal blocks (coupling $M$ and $A$) of $B$ to be approximately 0.
	
	While the three-qubit construction can be generalized to \textit{N}-qubit systems, only need to insert a pair of \textit{X} anyons for each additional qubit, as suggested in Fig.\ref{fig.8} (c), the exponential growth in matrix dimensionality renders the determination of EBMs computationally intractable. A critical challenge for future research lies in developing methodologies to obtain the exact forms of these high-dimensional EBMs.
	
	The braiding operation that form the standard gates for $N$-qubits include some that can be decomposed as direct products (e.g., $\sigma^{(6)}_1$can be decomposed as the direct product of $\sigma^{(3)}_1$ and the identity matrix, thereby linking two-qubit and one-qubit standard gates). Tensor product decomposition allows one to connect multi-qubit systems to systems with a smaller number of qubits, such as one- or two-qubit.
	
	\section{Conclusions}
	
	We investigate the metaplectic anyon models based on $SO(3)_2$ and obtain 6 anyon models according to \textit{F}-matrices/\textit{R}-symbols and fusion rules by non-standard encoding. By adding a pair of \textit{Z} anyons after braiding \textit{X} with $X^\prime$, the initial arrangement of anyons can be reset, which makes it feasible to obtain the corresponding EBMs. We emphasize that only auxiliary anyons are added for fusion in this process, which means that measurements may not be necessary to implement a universal set of gates for $SO(3)_2$ anyon model.The models of EBMs with global phase differences are grouped into one class, we select the best three models $\{V^{113}_3,V^{131}_3,V^{133}_1\}$ for systematic study. Utilizing \textit{F}-matrices and \textit{R}-symbols, we solve analytically the corresponding one-/two-qubit EBMs of these three models. For the one-qubit case, we use the global phase invariant distance as a metric, and based on the one-qubit EBMs, we use GA-enhanced SKA to construct a standard \textit{H}/\textit{T}-gate, then compare it with the Fibonacci model. The calculation results show that $V^{131}_3$ gives the best performance, although $V^{113}_3$ and $V^{133}_1$ are slightly inferior to the Fibonacci model at 0- and 1-level approximations, they are comparable to the Fibonacci model at 2- or 3-level  approximations. For the two-qubit case, we try to approximate a local equivalence class of the CNOT-gate. Based on the two-qubit EBMs, we use exhaustive search/GA to get the ideal braidword when the length is short/long. Compared with the Fibonacci model, the calculation results show that our model has obvious advantages, $\{V^{113}_3,V^{131}_3,V^{133}_1\}$ can obtain far higher precision (below $10^{-32}$). All three models with inverse EBMs find a braidword with $l$ = 20 and local equivalence class with distance equals 0. At the same time, the unitary measurement of the computational matrix always is better than the well-known Fibonacci model. We conducted a comparison between conventional and unconventional encoding schemes based on metaplectic anyons. The explicit forms of EBMs under conventional encoding were solved exactly, these EBMs satisfy the Artin braid group relations. However, attempts to construct a universal quantum gate set $\{H, T, CNOT\}$ using these EBMs failed. In contrast, after solving for the EBMs under unconventional encoding, these EBMs successfully constructed the universal gate set $\{H, T, CNOT\}$ with low error rates. Nevertheless, because this approach requires introducing additional $Z$-anyon for fusion, the continuous worldlines formed by anyon braiding become truncated, thus violating the Artin relations. Finally, we try to generalize the metaplectic anyon model to \textit{N}-qubits, but the exact solution is difficult due to the matrix dimension is too large. Our work increases the number of topological quantum computing models based on metaplectic anyons and is expected to be applied to future topological quantum computing.
	
	\textbf{Acknowledgments} This work was supported by the Shanghai Science and Technology Innovation Action Plan (Grant No. 24LZ1400800), National Natural Science Foundation of China (Grant Nos. 12374046, 11204261), College of Physics and Optoelectronic Engineering training program, a Key Project of the Education Department of Hunan Province (Grant No. 19A471), Natural Science Foundation of Hunan Province (Grant No. 2018JJ2381). 
	
	\textbf{Data availability} The data that support the findings of this article are not publicly available. The data are available from the authors upon reasonable request.
	
	\appendix
	
	\begin{widetext}
		
		\section{The process of selecting available models}
		
		According to the fusion rules, qubits can be constructed requiring that the fusion result of the intermediate state have two kinds of anyons corresponding to the $\ket{0}$ and $\ket{1}$.
		The following fusion rules must be applied:
		
		\begin{equation}
			\small
			\begin{split}
				X\otimes X=1\oplus Y,X\otimes Y=X\oplus X^\prime,
				X\otimes X^\prime=Y\oplus Z,
				Y\otimes X^\prime=X\oplus X^\prime,X^\prime\otimes X^\prime=1\oplus Y.
			\end{split}
		\end{equation}
		
		Three anyons construct qubit, the first and second anyons correspond to the two anyons of the above fusion rule, and the third anyon is selected from $\{1,X,Y,X^\prime,Z\}$ to obtain the following model:
		
		\begin{center}
			$V^{111}_1,V^{112}_2,V^{113}_3,V^{331}_1,V^{332}_2,V^{333}_3,V^{121}_2,$
			$V^{122}_1,V^{122}_3,V^{123}_2,V^{211}_2,V^{212}_1,V^{212}_3,V^{213}_2,$\\
			$V^{131}_3,V^{132}_2,V^{133}_1,V^{311}_3,V^{312}_2,V^{313}_1,V^{231}_2,$
			$V^{232}_1,V^{232}_3,V^{233}_2,V^{321}_2,V^{322}_1,V^{322}_3,V^{323}_2.$
		\end{center}
		
		There are 28 possible models in total. With the exception of the standard encodings $V^{111}_1$ and $V^{333}_3$, non-standard encodings are unusable due to state changes (changes in anyons initial order) after a single exchange. In the model of adjacent different kinds of anyons, we make the model with only 1 and 3 adjacent can be used by introducing a pair of \textit{Z} anyons after braiding for once as show in Fig.\ref{fig.9}.
		
		\begin{figure}[h]
			\centering
			\includegraphics[width=0.5\textwidth]{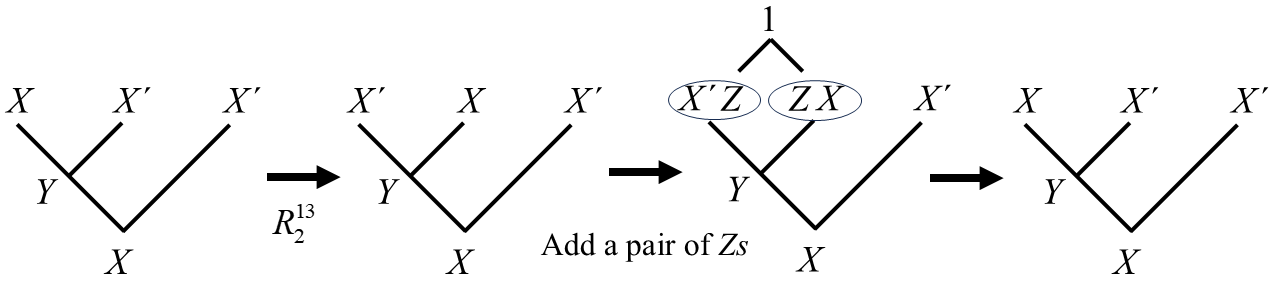}
			\caption{The process of introducing a pair of $Z$   fused with $X$ and $X^{\prime}$ individually after braiding $X$ with $X^{\prime}$ to restore the original anyonic ordering.}
			\label{fig.9}
		\end{figure}
		
		The fusion rule with \textit{Z} is Abelian, the fusion of the other anyons with \textit{Z} produces the only one anyon. we notice that $X\otimes Z=X^\prime$ and $X^\prime\otimes Z=X$, means \textit{X} switches with $X^\prime$, which is exactly what we want. But this is not true for \textit{X} adjacent to \textit{Y} or $X^\prime$ adjacent to \textit{Y}. The only thing we have to worry about is whether the introduction of a pair of \textit{Z}-anyons will have any effect on the fusion result, fortunately it won't, we can understand this  process through Fig.\ref{fig.10}.
		
		\begin{figure}[h]
			\centering
			\includegraphics[width=0.5\textwidth]{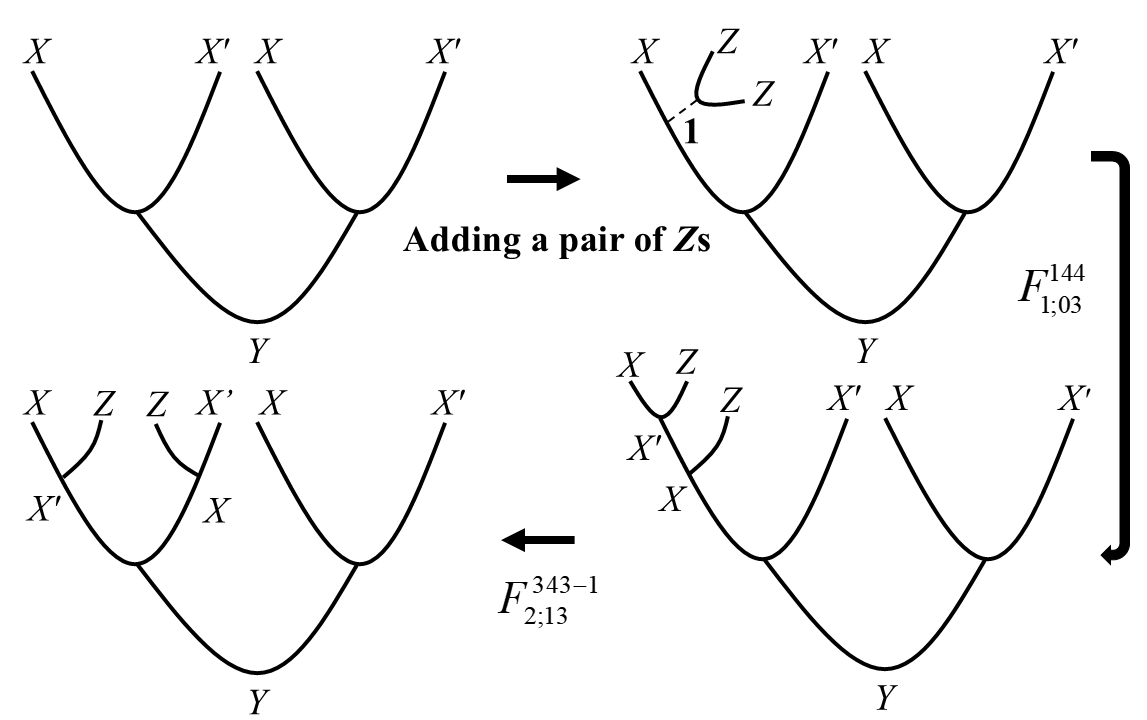}
			\caption{Producing a pair of $Z$-type anyons from vacuum followed by individual fusion with $X$ and $X^{\prime}$ anyons, mediating their exchange via two successive trivial $F$-operations.}
			\label{fig.10}
		\end{figure}
		
		\textit{Y} doesn't convert to \textit{Z} in this process.
		
		Only the following models can be used:
		
		\begin{center}
			$V^{111}_1,V^{333}_3,V^{113}_3,V^{331}_1,V^{131}_3,V^{133}_1,V^{311}_3,V^{313}_1.$
		\end{center}
		
		If we brading \textit{X} and $X^\prime$, we can return to the original state by introducing a pair of \textit{Z} anyons, and braiding the same kind of anyons requires no additional processing.The $\sigma_1^{(3)}$ and $\sigma_2^{(3)}$ of all the above models are solved and the corresponding EMBs $\{\sigma_1^{(6)},\sigma_2^{(6)},\sigma_4^{(6)},\sigma_5^{(6)}\}$ of local equivalent class of two-qubit gates can also be obtained easily by direct sum and direct product. Such as: 
		
		\begin{center}
			$\sigma_1^{(6)}=R\oplus(\sigma_1^{(3)}\otimes I_2),
			\sigma_2^{(6)}=R\oplus(\sigma_2^{(3)}\otimes I_2),$\\
			$\sigma_4^{(6)}=R\oplus(I_2\otimes \sigma_1^{(3)}),
			\sigma_5^{(6)}=R\oplus(I_2\otimes  \sigma_2^{(3)}) $.
		\end{center}
		
		Only $\sigma_3^{(6)}$ must be obtained by processing of braiding the third and fourth anyons. Then we can construct one-qubit \textit{H}-/\textit{T}-gates and two-qubit local equivalent class [CNOT] using these EBMs, and we find $V_1^{111}$ and $V_3^{333}$ construct \textit{H}-/\textit{T}-gates failure.
		
		We put together the remaining 6 models in Table \ref{tab.3} which the one-qubit EBMs differ only by a global phase:
		
		\begin{table}[h]
			\centering
			\caption{The 6 models of three anyons based on $SO(3)_2$.}
			\begin{tabular}{lc}
				\hline
				\hline
				Model&Phase difference\\
				\hline
				$V^{113}_3/V^{331}_1$&$\sigma_1^{(3)}(\pi)$\\
				$V^{131}_3/V^{313}_1$&same\\
				$V^{311}_3/V^{133}_1$&$\sigma_2^{(3)}(\pi)$\\
				\hline
				\hline
			\end{tabular}
			\label{tab.3}
		\end{table}

		The phase difference between the first model and the second model in the summarized in parentheses. 
		
		For one-qubit, the same braidword formed by a set of EBMs that differ by a global phase has the same result calculated using the global phase invariant distance formula. However, for two-qubit, the same braidword is constructed by two sets of EBMs with a global phase difference, and the distance from CNOT-gate is calculated using local equivalence class, and different results are obtained.
		
		We use an exhaustive search to find the minimum value of local equivalence class [CNOT] distance at each length for these six models. The result are presented in Table \ref{tab.4} and Table \ref{tab.5}.
		
		\begin{table}[h]
			\centering
			\caption{The minimum distances of local equivalence class [CNOT]. The inverse matrices are not added.}
			\begin{tabular}{ccccccc}
				\hline
				\hline
				Length&$V^{113}_3$&$V^{331}_1$ &$V^{131}_3$&$V^{313}_1$&$V^{311}_3$&$V^{133}_1$\\
				\hline
				3&5&5&5&5&5&5\\
				4&5&5&5&5&5&5\\
				5&5&5&5&5&5&5\\
				6&5&5&5&5&5&5\\
				7&$1.28\times 10^{-33}$&5&$2.70\times 10^{-35}$&$1.83\times 10^{-32}$&5&5\\
				8&$1.23\times 10^{-32}$&5&$1.23\times 10^{-32}$&$1.23\times 10^{-32}$&5&5\\
				9&$1.23\times 10^{-32}$&5&$1.23\times 10^{-32}$&$1.23\times 10^{-32}$&5&5\\
				10&$9.46\times 10^{-63}$&1.79&$1.16\times 10^{-62}$&$2.57\times 10^{-62}$&$2.48\times 10^{-32}$&$1.98\times 10^{-31}$\\
				11&$1.23\times 10^{-32}$&$3.11\times 10^{-3}$&$1.23\times 10^{-32}$&$1.23\times 10^{-32}$&$1.23\times 10^{-32}$&$1.23\times 10^{-32}$\\
				12&$3.26\times 10^{-35}$&$9.80\times 10^{-6}$&$3.08\times 10^{-36}$&$3.24\times 10^{-34}$&$1.23\times 10^{-32}$&$1.23\times 10^{-32}$\\
				13&$3.75\times 10^{-43}$&$2.36\times 10^{-8}$&$1.24\times 10^{-43}$&$2.38\times 10^{-43}$&$7.24\times 10^{-37}$&$4.00\times 10^{-38}$\\
				\hline
				\hline
			\end{tabular}
			\label{tab.4}
		\end{table}
		
		\begin{table}[h]
			\centering
			\caption{The minimum distances of local equivalence class [CNOT]. The inverse matrices are added.}
			\begin{tabular}{ccccccc}
				\hline
				\hline
				Length&$V^{113}_3$&$V^{331}_1$&$V^{131}_3$&$V^{313}_1$&$V^{311}_3$&$V^{133}_1$\\
				\hline
				3&5&5&5&5&5&5\\
				4&5&5&5&5&5&5\\
				5&5&5&5&5&5&5\\
				6&$1.23\times 10^{-32}$&5&5&$5.00\times 10^{-5}$&5&5\\
				7&$2.37\times 10^{-37}$&4.40&$2.70\times 10^{-35}$&$1.23\times 10^{-32}$&5&5\\
				\hline
				\hline
			\end{tabular}
			\label{tab.5}
		\end{table}
		We can find that $V^{113}_3$ was significantly better than $V^{331}_1$, and the other two groups were not much different. We choose $\{V^{113}_3,V^{131}_3,V^{133}_1\}$ for the study of constructing a standard quantum gate systematically.	
		
		\section{The all of \textit{F}-matrices and \textit{R}-symbols used to solve BEMs}
		
		All local variations are based on the element of \textit{F}-matrix and \textit{R}-symbol, as defined in the Fig.\ref{fig.11}.
		
		\begin{figure}[h]
			\centering
			\includegraphics[width=0.5\textwidth]{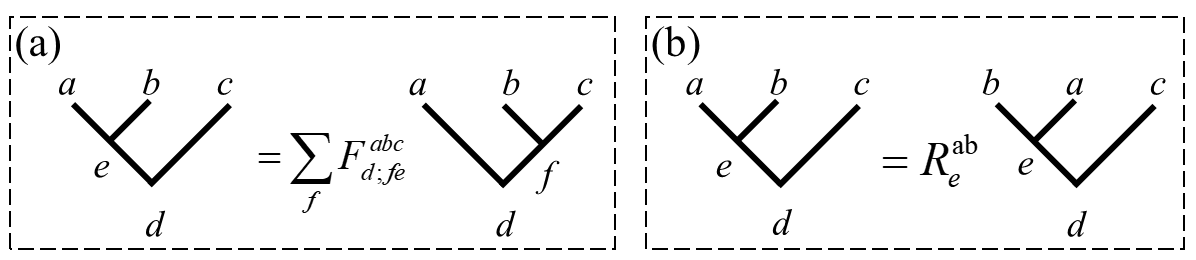}
			\caption{(a) Definition of $F^{abc}_{d;fe}$. (b) Definition of $R^{ab}_e$.}
			\label{fig.11}
		\end{figure}
		
		We list the \textit{F}-matrices and \textit{R}-symbols for $\{V^{113}_3,V^{131}_3,V^{133}_1,V^{111}_1\}$ as follows:\\
		
		\begin{center}
			$F^{111}_1=
			\begin{bmatrix}
				F^{111}_{1;00}&F^{111}_{1;02}\\
				F^{111}_{1;20}&F^{111}_{1;22}
			\end{bmatrix}=\frac{1}{\sqrt{3}}
			\begin{bmatrix}
				-1&\sqrt{2}\\
				\sqrt{2}&1
			\end{bmatrix},$
			$F^{113}_3=
			\begin{bmatrix}
				F^{113}_{3;20}&F^{113}_{3;22}\\
				F^{113}_{3;40}&F^{113}_{3;42}
			\end{bmatrix}=\frac{1}{\sqrt{3}}
			\begin{bmatrix}
				-\sqrt{2}&1\\
				1&\sqrt{2}
			\end{bmatrix},$
			$F^{311}_3=
			\begin{bmatrix}
				F^{311}_{3;02}&F^{311}_{3;04}\\
				F^{311}_{3;22}&F^{311}_{3;24}
			\end{bmatrix}=\frac{1}{\sqrt{3}}
			\begin{bmatrix}
				-\sqrt{2}&1\\
				1&\sqrt{2}
			\end{bmatrix},$
			$F^{112}_2=
			\begin{bmatrix}
				F^{112}_{2;10}&F^{112}_{2;12}\\
				F^{112}_{2;30}&F^{112}_{2;32}
			\end{bmatrix}=\frac{1}{\sqrt{2}}
			\begin{bmatrix}
				-1&1\\
				1&1
			\end{bmatrix},$\\
			$F^{131}_3=
			\begin{bmatrix}
				F^{131}_{3;22}&F^{131}_{3;24}\\
				F^{131}_{3;42}&F^{131}_{3;44}
			\end{bmatrix}=\frac{1}{\sqrt{3}}
			\begin{bmatrix}
				-1&\sqrt{2}\\
				\sqrt{2}&1
			\end{bmatrix},$
			$F^{133}_1=
			\begin{bmatrix}
				F^{133}_{1;02}&F^{133}_{1;04}\\
				F^{133}_{1;22}&F^{133}_{1;24}
			\end{bmatrix}=\frac{1}{\sqrt{3}}
			\begin{bmatrix}
				-\sqrt{2}&1\\
				1&\sqrt{2}
			\end{bmatrix},$
			$F^{331}_1=
			\begin{bmatrix}
				F^{331}_{1;20}&F^{331}_{1;22}\\
				F^{331}_{1;40}&F^{331}_{1;42}
			\end{bmatrix}=\frac{1}{\sqrt{3}}
			\begin{bmatrix}
				-\sqrt{2}&1\\
				1&\sqrt{2}
			\end{bmatrix},$
			$F^{332}_2=
			\begin{bmatrix}
				F^{332}_{2;10}&F^{332}_{2;12}\\
				F^{332}_{2;30}&F^{332}_{2;32}
			\end{bmatrix}=\frac{1}{\sqrt{2}}
			\begin{bmatrix}
				1&-1\\
				-1&\-1
			\end{bmatrix},$
		\end{center}
		
		\begin{center}
			$R^{11}_0 =e^{i\frac{3\pi}{4}},
			R^{11}_2 =e^{i\frac{\pi}{12}},
			R^{13}_2 =R^{31}_2 =e^{i\frac{7\pi}{12}},
			R^{13}_4 =R^{31}_4 =e^{i\frac{\pi}{4}},
			R^{33}_0 =e^{-i\frac{\pi}{4}},
			R^{33}_2 =e^{-i\frac{11\pi}{12}}.$
		\end{center}
		
		\section{The process of precisely solving the EBMs}
		
		Note that in the calculation processes, we ignore all \textit{F}-matrices whose fusion result is vacuum, because it is always trivial.
		
		\begin{figure}[h]
			\centering
			\includegraphics[width=0.8\textwidth]{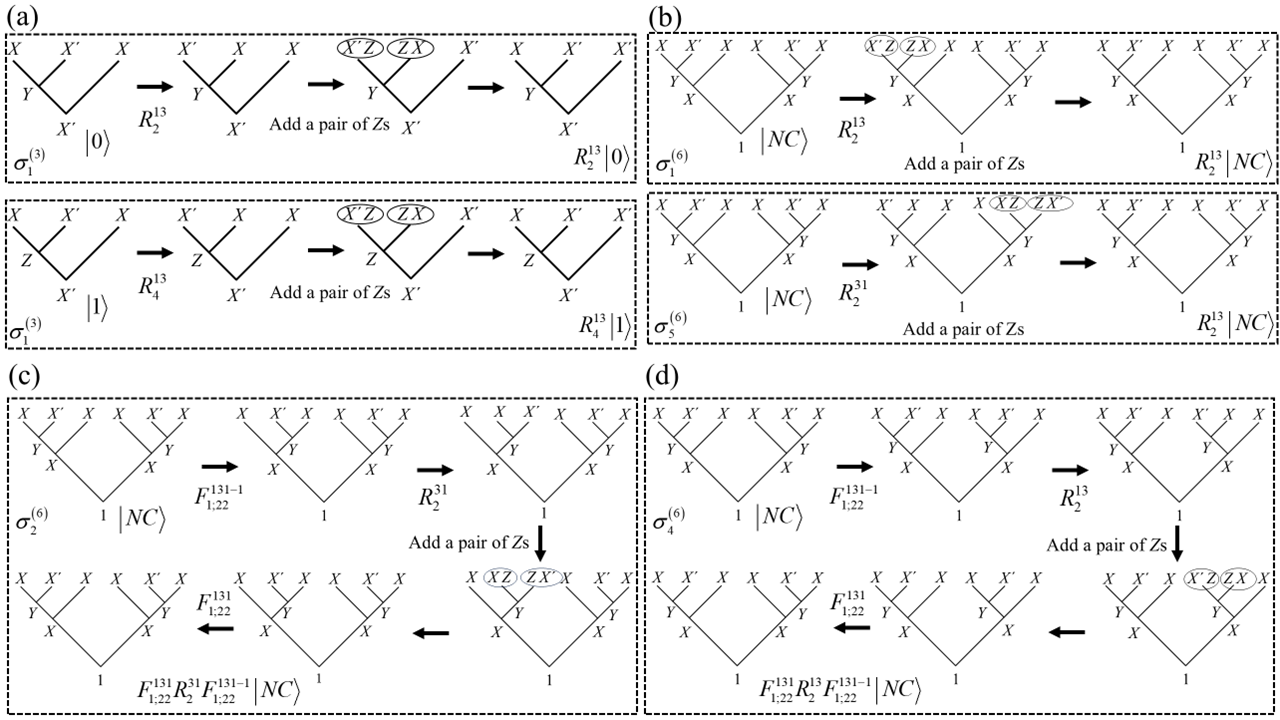}
			\caption{Braiding Operations in the $V^{131}_3$ Model. (a) One-qubit case: The process of using $\sigma^{(3)}_1$ operations acting on $\ket{0}$ and $\ket{1}$ respectively to make $\ket{0}$ and $\ket{1}$ acquire phase. (b) Two-qubit case: The process of using $\sigma^{(6)}_1$ and $\sigma^{(6)}_5$ operations acting on $\ket{NC}$ respectively to make $\ket{NC}$ acquire phase. (c) Two-qubit case: The process of using $\sigma^{(6)}_2$ operation acting on $\ket{NC}$ to make $\ket{NC}$ acquire phase. (d) Two-qubit case: The process of using $\sigma^{(6)}_4$ operation acting on $\ket{NC}$ to make $\ket{NC}$ acquire phase.}
			\label{fig.12}
		\end{figure}
		
		Taking Model $V^{131}_3$ as an example, the Fig.\ref{fig.12} illustrates the braiding process for solving EBMs. Fig.\ref{fig.12} (a) depicts the action of $\sigma^{(3)}_1$ (braiding the first anyon \textit{X} and the second anyon $X^\prime$ on $\ket{0}$ and $\ket{1}$. After braiding, a pair of \textit{Z}-anyons is inserted to restore the initial anyon order, yielding $R^{13}_2\ket{0}$ and $R^{13}_4\ket{1}$ finally. By substituting the \textit{R}-symbol values from Appendix B, the explicit form of the braiding matrix $\sigma^{(3)}_1$ is systematically derived. Fig.\ref{fig.12} (b) corresponds to the braiding process of $\sigma^{(6)}_1\slash\sigma^{(6)}_5$ acting on $\ket{NC}$, while Fig.\ref{fig.12} (c)/(d) show$\sigma^{(6)}_2\slash\sigma^{(6)}_3$ acting on $\ket{NC}$. Similarly, by applying the operations of \textit{F} and \textit{R}  to $\sigma^{(6)}_1\slash\sigma^{(6)}_2\slash\sigma^{(6)}_4\slash\sigma^{(6)}_5$ acting on $\ket{00}$, $\ket{01}$, $\ket{10}$ and $\ket{11}$ respectively, then corresponding braiding matrices are constructed by inserting the results into a matrix.
		
		We present an example of the braiding process for $\sigma_2^{(3)}/\sigma_3^{(6)}$ acting on $\ket{0}$ or $\ket{00}$ by the figure and other calculation results by formula.
		
		Calculating $\sigma_2^{(3)}$ of the $\{V^{113}_3,V^{131}_3,V^{133}_1,V^{111}_1\}$ model:
		
		\begin{figure}[h]
			\centering
			\includegraphics[width=0.5\textwidth]{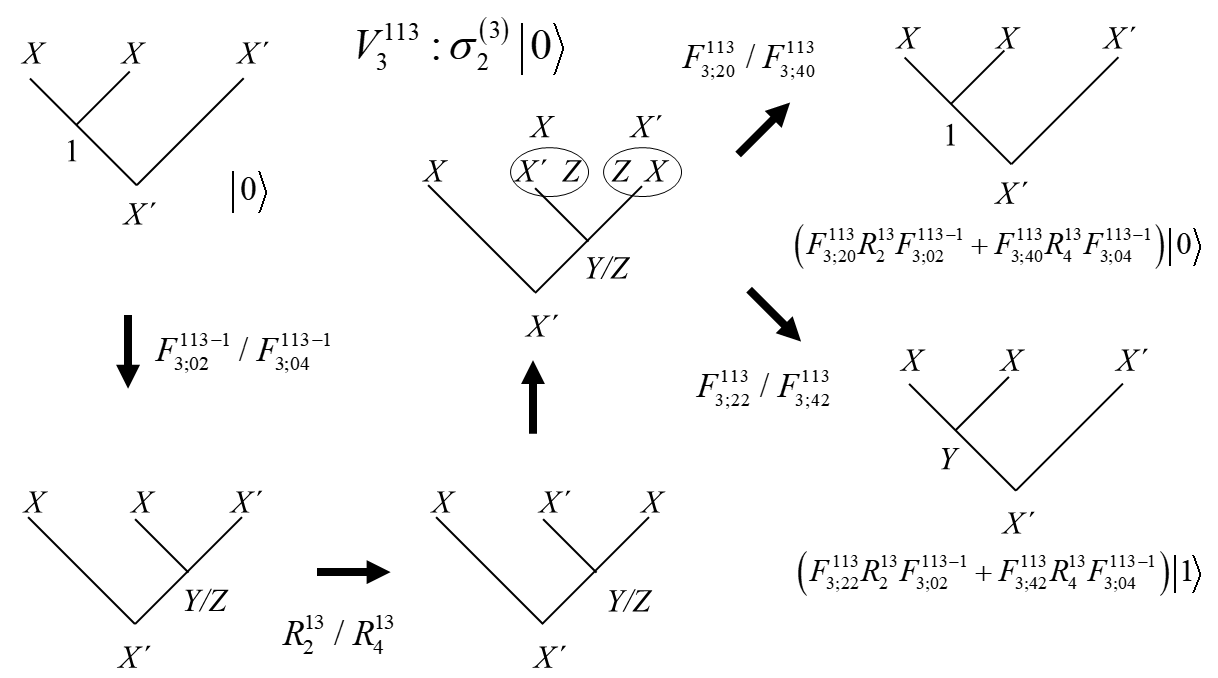}
			\caption{Braiding Operations in the $V^{113}_3$ Model, the process of $\sigma^{(3)}_2$ act on $\ket{0}$.}
			\label{fig.13}
		\end{figure}
		
		To clarify the aforementioned process, we provide the following elucidation. As established in prior discussions, $((X,X)_1,X^\prime)_{X^\prime}$ is defined as $\ket{0}$. The operation $\sigma_2^{(3)}$ denoted as braiding the second and third anyons. The fusion basis undergoes transformation via the \textit{F}-matrix, during which intermediate states evolve from $((X,X)_1,X^\prime)_{X^\prime}$ to $(X,(X,X^\prime)_Y)_{X^\prime}$ and $((X,(X,X^\prime)_Z)_{X^\prime}$. Notably, two distinct fusion channels coexist in this framework, they are put together for diagrammatic conciseness.Subsequently, the \textit{R}-matrix governs the braiding of $X^\prime$ and $X$, imparting a topological phase to the resultant state. This operation inverts the anyon ordering to $X$ and $X^\prime$, necessitating the introduction of a $Z$-anyon pair to restore the original topological charge sequence. A final \textit{F}-matrix operation reverts the system to its initial fusion configuration. Owing to fusion rules, the terminal state $\ket{0}$ and $\ket{1}$ manifests as a superposition of the computational basis states, with coefficients dictated by the elements of \textit{F}-matrix and \textit{R}-symbol of the braiding process. The corresponding computational protocol aligns with the $\sigma^{(3)}_2\ket{0}$ formalism in the $V^{113}_3$, the corresponding braiding process is shown in Fig.\ref{fig.13}. As expressed in the equations below:
		
		$V^{113}_3:$\\
		\begin{center}
			$\sigma^{(3)}_2\ket{0}=(F^{113}_{3;20}R^{13}_2F^{113-1}_{3;02}+F^{113}_{3;40}R^{13}_4F^{113-1}_{3;04})\ket{0}+(F^{113}_{3;22}R^{13}_2F^{113-1}_{3;02}+F^{113}_{3;42}R^{13}_4F^{113-1}_{3;04})\ket{1},$\\
			$\sigma^{(3)}_2\ket{1}=(F^{113}_{3;20}R^{13}_2F^{113-1}_{3;22}+F^{113}_{3;40}R^{13}_4F^{113-1}_{3;24})\ket{0}+(F^{113}_{3;22}R^{13}_2F^{113-1}_{3;22}+F^{113}_{3;42}R^{13}_4F^{113-1}_{3;24})\ket{1}.$
		\end{center}
		
		By applying the braiding operator $\sigma^{(3)}_2$ to the computational basis states $\ket{0}$ and $\ket{1}$, and populating a two-dimensional matrix with coefficients derived from the corresponding elements of \textit{F}-matrix and \textit{R}-symbol, we obtain the EBM $\sigma^{(3)}_2$. This methodology can be generalized to systematically construct EBMs for other anyon models as bellow.
		
		$V^{131}_3:$\\
		\begin{center}
			$\sigma^{(3)}_2\ket{0}=(F^{131}_{3;22}R^{13}_2F^{131-1}_{1;22}+F^{131}_{1;42}R^{13}_4F^{131-1}_{1;24})\ket{0}+(F^{131}_{1;24}R^{13}_2F^{131-1}_{1;22}+F^{131}_{1;44}R^{13}_4F^{131-1}_{1;24})\ket{1},$\\
			$\sigma^{(3)}_2\ket{1}=(F^{131}_{1;22}R^{13}_2F^{131-1}_{1;42}+F^{131}_{1;42}R^{13}_4F^{131-1}_{1;44})\ket{0}+(F^{131}_{1;24}R^{13}_2F^{131-1}_{1;42}+F^{131}_{1;44}R^{13}_4F^{131-1}_{1;44})\ket{1}.$
		\end{center}
		
		$V^{133}_1:$\\
		\begin{center}
			$\sigma^{(3)}_2\ket{0}=(F^{133}_{1;02}R^{33}_0F^{133-1}_{1;20}+F^{133}_{1;22}R^{33}_2F^{133-1}_{1;22}\ket{0}+(F^{133}_{1;04}R^{33}_0F^{133-1}_{1;20}+F^{133}_{1;24}R^{33}_2F^{133-1}_{1;22}\ket{1},$\\
			$\sigma^{(3)}_2\ket{0}=(F^{133}_{1;02}R^{33}_0F^{133-1}_{1;40}+F^{133}_{1;22}R^{33}_2F^{133-1}_{1;42}\ket{0}+(F^{133}_{1;04}R^{33}_0F^{133-1}_{1;40}+F^{133}_{1;24}R^{33}_2F^{133-1}_{1;42}\ket{1}.$	
		\end{center}
		
		$V^{111}_1:$\\
		\begin{center}
			$\sigma^{(3)}_2\ket{0}=(F^{111}_{1;00}R^{11}_0F^{111-1}_{1;00}+F^{111}_{1;20}R^{11}_2F^{111-1}_{1;02}\ket{0}+(F^{111}_{1;02}R^{11}_0F^{111-1}_{1;00}+F^{111}_{1;22}R^{11}_2F^{111-1}_{1;02}\ket{1},$\\
			$\sigma^{(3)}_2\ket{0}=(F^{111}_{1;00}R^{11}_0F^{111-1}_{1;20}+F^{111}_{1;20}R^{11}_2F^{111-1}_{1;22}\ket{0}+(F^{111}_{1;02}R^{11}_0F^{111-1}_{1;20}+F^{111}_{1;22}R^{11}_2F^{111-1}_{1;22}\ket{1}.$	
		\end{center}

		Calculating $\sigma_3^{(6)}$ of the $\{V^{113}_3,V^{131}_3,V^{133}_1,V^{111}_1\}$ model:
		
		\begin{figure}[h]
			\centering
			\includegraphics[width=0.5\textwidth]{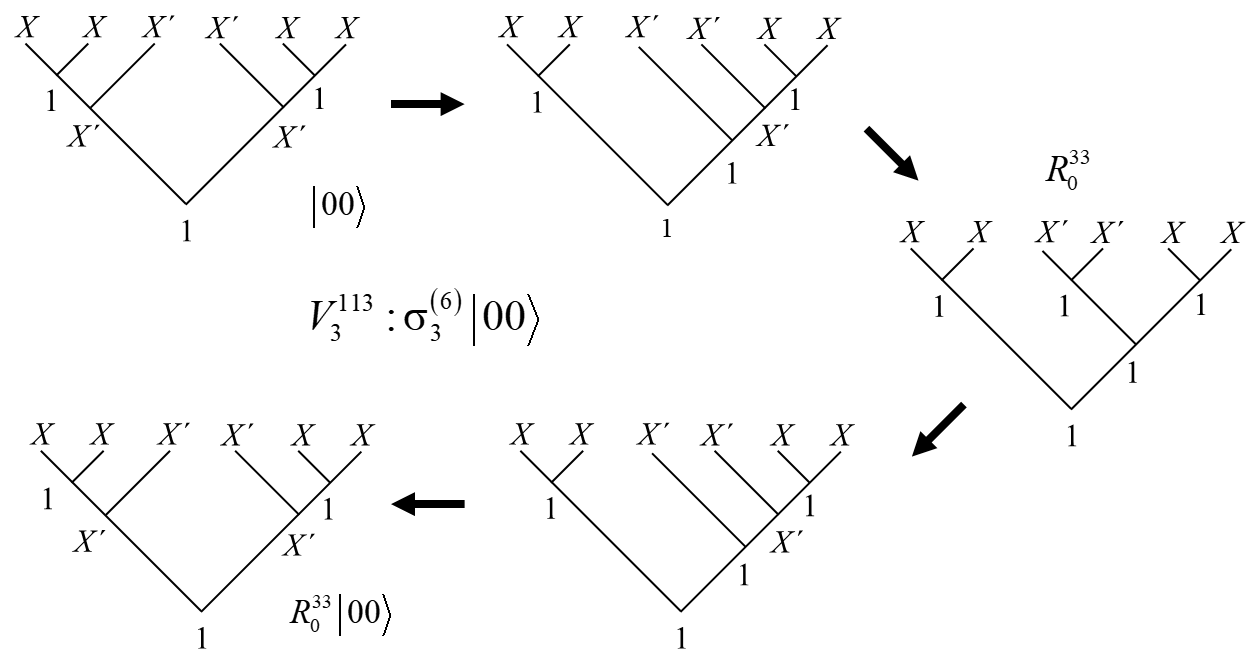}
			\caption{Braiding Operations in the $V^{113}_3$ Model, the process of $\sigma^{(6)}_3$ act on $\ket{00}$.}
			\label{fig.14}
		\end{figure}
		
		The process illustrated in the Fig.\ref{fig.14} demonstrates the action of $\sigma_3^{(6)}$ on the $\ket{00}$ state in the 
		$\{V^{113}_3\}$ model. Throughout this braiding protocol, the 
		\textit{F}-matrix remains trivial (yielding a single fusion outcome) and is therefore omitted. Furthermore, since the braiding operation involves identical anyons, no auxiliary 
		\textit{Z}s are required. The braiding effect on this state is governed solely by the phase factor introduced through the \textit{R}-symbol.
		
		The methodology for determining the EBMs in the two-qubit system parallels the single-qubit case. However, the two-qubit Hilbert space contains four computational basis states $\{\ket{00},\ket{01},\ket{10},\ket{11}\}$, and one non-computational basis state. To construct the full braiding matrix, we systematically apply $\sigma^{(6)}$ to each computational and non-computational state, then populate a five-dimensional matrix with the corresponding coefficients of state. The complete derivation is as detailed in the following equations.
		
		$V^{113}_3:$
		\begin{center}
			$\sigma^{(6)}_3\ket{NC}=(F^{332-1}_{2;01}R^{33}_0F^{332}_{2;10}+F^{332-1}_{2;21}R^{33}_2F^{332}_{2;12})\ket{NC}+(F^{332-1}_{2;03}R^{33}_2F^{332}_{2;10}+(F^{332-1}_{2;23}R^{33}_0F^{332}_{2;12})\ket{11},$
			$\sigma^{(6)}_3\ket{00}=R^{33}_0\ket{00},
			\sigma^{(6)}_3\ket{01}=R^{33}_2\ket{01},
			\sigma^{(6)}_3\ket{10}=R^{33}_2\ket{10},$
			$\sigma^{(6)}_3\ket{11}=(F^{332-1}_{2;01}R^{33}_0F^{332}_{2;30}+F^{332-1}_{2;21}R^{33}_2F^{332}_{2;32})\ket{NC}+(F^{332-1}_{2;03}R^{33}_0F^{332}_{2;30}+(F^{332-1}_{2;23}R^{33}_2F^{332}_{2;32})\ket{11}.$
		\end{center}
		
		$V^{131}_3:$	
		\begin{center}
			$\sigma^{(6)}_3\ket{NC}=(F^{112-1}_{2;01}R^{11}_0F^{112}_{2;10}+F^{112-1}_{2;21}R^{11}_2F^{112}_{2;12})\ket{NC}+(F^{112-1}_{2;03}R^{11}_0F^{112}_{2;10}+F^{112-1}_{2;23}R^{11}_2F^{112}_{2;12})\ket{00},$
			
			$\sigma^{(6)}_3\ket{00}=(F^{112-1}_{2;01}R^{11}_0F^{112}_{2;30}+F^{112-1}_{2;21}R^{11}_2F^{112}_{2;32})\ket{NC}+(F^{112-1}_{2;03}R^{11}_0F^{112}_{2;30}+F^{112-1}_{2;23}R^{11}_2F^{112}_{2;32})\ket{00},$
			
			$\sigma^{(6)}_3\ket{01}=R^{11}_2\ket{01},
			\sigma^{(6)}_3\ket{10}=R^{11}_2\ket{10},
			\sigma^{(6)}_3\ket{11}=R^{11}_0\ket{11}.$
		\end{center}
		
		$V^{133}_1:$	
		\begin{center}
			$\sigma^{(6)}_3\ket{NC}=(F^{332-1}_{2;03}R^{33}_0F^{332}_{2;30}+F^{332-1}_{2;23}R^{33}_2F^{332}_{2;32})\ket{NC}+(F^{332-1}_{2;01}R^{33}_0F^{332}_{2;30}+F^{332-1}_{2;21}R^{33}_2F^{332}_{2;32})\ket{00},$
			
			$\sigma^{(6)}_3\ket{00}=(F^{332-1}_{2;03}R^{33}_0F^{332}_{2;10}+F^{332-1}_{2;23}R^{33}_2F^{332}_{2;12})\ket{NC}+(F^{332-1}_{2;01}R^{33}_0F^{332}_{2;10}+F^{332-1}_{2;21}R^{33}_2F^{332}_{2;12})\ket{00},$
			
			$\sigma^{(6)}_3\ket{01}=R^{33}_2\ket{01},
			\sigma^{(6)}_3\ket{10}=R^{33}_2\ket{10},
			\sigma^{(6)}_3\ket{11}=R^{33}_0\ket{11}.$
		\end{center}
		
		$V^{111}_1:$	
		\begin{center}
			$\sigma^{(6)}_3\ket{NC}=(F^{112-1}_{2;03}R^{11}_0F^{112}_{2;30}+F^{112-1}_{2;23}R^{11}_2F^{112}_{2;32})\ket{NC}+(F^{112-1}_{2;01}R^{11}_0F^{112}_{2;30}+F^{112-1}_{2;21}R^{11}_2F^{112}_{2;32})\ket{00},$
			
			$\sigma^{(6)}_3\ket{00}=R^{11}_0\ket{00},
			\sigma^{(6)}_3\ket{01}=R^{11}_2\ket{01},
			\sigma^{(6)}_3\ket{10}=R^{11}_1\ket{10}.$
			
			$\sigma^{(6)}_3\ket{11}=(F^{112-1}_{2;03}R^{11}_0F^{112}_{2;10}+F^{112-1}_{2;23}R^{11}_2F^{112}_{2;12})\ket{NC}+(F^{112-1}_{2;01}R^{11}_0F^{112}_{2;10}+F^{112-1}_{2;21}R^{11}_2F^{112}_{2;12})\ket{00},$
		\end{center}	
	\end{widetext}
	
	\begin{widetext}
		\section{GA-enhanced SKA}
		We provide a brief overview of GA-enhanced SKA for quantum compilation problems below. For implementation details, see the references \cite{RN38}.
		
		Specific practices of GA are as follows:
		
		(\romannumeral1) \{A braidword / many braidwords / global phase invariant distance (Eq .\eqref{eq.3})\} corresponds to \{an individual / a population / fitness\}.
		
		(\romannumeral2) To randomly generate a number of braidwords, to produce a new generation of population through hybridization and mutation as the parent’s text.
		
		(\romannumeral3) Retaining some individuals with high fitness (low global phase invariant distance (Eq .\eqref{eq.3})) in the parent's text.
		
		(\romannumeral4) A new generation of population is generated by hybridization and mutation of the parent’s text while retaining a part of the population with high fitness as the new parent’s text.
		
		(\romannumeral5) Repeat steps (\romannumeral3) and (\romannumeral4) until reach the hybridization generation we set.
		
		(\romannumeral6) The individual with the highest final output fitness.
		
		Definition of Key Algorithmic Terms:
		
		1. Individual: A matrix formed by multiplying a sequence of specified EBMs. For example, the braidword ABAD corresponds to the matrix resulting from the product $\sigma_1$$\sigma_2$$\sigma_1$$\sigma^{-1}_2$.
		
		2. Population: A collection of individuals, each represented by a distinct braidword.
		
		3. Crossover: An operation where two braidwords are randomly selected from the population and combined to generate two new braidwords. For instance, given braid words CBABC and ADBDD, a random integer (e.g., 3) is selected as the crossover point. The resulting offspring are CBADD and ADBBC.
		
		4. Mutation: Following crossover, newly generated braidwords undergo mutation with a defined probability. If mutation occurs, a random position within the braidword is selected (e.g., position 2). The EBM at this position is replaced by another valid EBM (e.g., B in CBADD could be replaced by A/C/D).
		
		5. Hybridization Generation: The total number (G) of offspring populations generated through successive hybridization operations, starting from the initial population (generation 0). Specifically, population 1 is generated by hybridizing population 0, population 2 is generated by hybridizing population 1, and so forth, until population G is generated by hybridizing population G-1.
		
		The designations A, B, C, and D correspond to the braid group generators $\sigma_1$, $\sigma_2$, $\sigma^{-1}_1$, and $\sigma^{-1}_2$, respectively.
		
		Specific practices of SKA are as follows:
		
		(\romannumeral1) 0-level Approximation: Determine a target one-qubit gate (e.g., Hadamard gate $H$) and a standard braid length $L_0$
		(number of EBMs). Obtain its 0-level approximation ($H_0$) via exhaustive search.
		
		(\romannumeral2) 1-level Approximation: Compute the deviation matrix ($\triangle_1 = H H^{\dagger}_0$), then perform GC-decomposition ($\triangle_1 = V W V^{\dagger} W^{\dagger}$). Find 0-level approximations $V_0$ and $W_0$ for matrices $V$ and $W$ (using exhaustive search). Construct the 1-level approximation finally ($H_1 = V_0 W_0 V_0^{\dagger} W_0^{\dagger} H_0$)
		
		(\romannumeral3) 2-level Approximation: Compute the new deviation matrix ($\triangle_2 = H H^{\dagger}_1$), then perform GC-decomposition ($\triangle_2 = V^{\prime} W^{\prime} (V^{\prime})^{\dagger} (W^{\prime})^{\dagger}$). Find 1-level approximations $V^{\prime}_1$ and $W^{\prime}_1$ for matrices $V^{\prime}$ and $W^{\prime}$ (Note: This requires applying the 1-level approximation procedure to $V^{\prime}$ and $W^{\prime}$ individually). Construct the 2-level approximation finally ($H_2 = V^{\prime}_1 W^{\prime}_1 (V^{\prime}_1)^{\dagger} (W^{\prime}_1)^{\dagger} H_1$)
		
		(\romannumeral4) The process extends recursively to n-level: Compute the $\triangle_n = H H^{\dagger}_{n-1}$,  perform decomposition $\triangle_n = V_n W_n (V_n)^{\dagger} (W_n^{\prime})^{\dagger}$). Recursively obtain (n-1)-level approximations $V_{n-1}$, $W_{n-1}$, $H_{n-1}$ for $V_n$, $W_n$, $H$ respectively. Construct the n-level approximation finally $H_n = V_{n-1} W_{n-1} (V_{n-1})^{\dagger} (W_{n-1})^{\dagger} H_{n-1}$
		
		The computational time increases by a factor of approximately three per recursion level, as each GC decomposition necessitates two additional exhaustive searches (for $V$ and $W$). The braid word length scales by a factor of five per level, as $H_n$ incorporates five components: $ V_{n-1}$, $W_{n-1}$, $(V_{n-1})^{\dagger}$, $(W_{n-1})^{\dagger}$, and $H_{n-1}$. The core of the SKA  lies in GC-decomposition, which relies on solving the equation $sin(\theta/2)=2sin^2(\phi/2)\sqrt{1-sin^4(\phi/2)}$. For further details regarding SKA, are referred to Reference \cite{RN13}.
		
		The integration of GA with SKA is implemented by replacing the exhaustive search required in the original SKA algorithm with GA. In each iteration, the optimal braidword identified by the GA serves as the 0-level approximation for the SKA procedure.
		
		It should be noted that while GA-enhanced SKA is employed for constructing one-qubit gates using metaplectic anyons, the SKA algorithm appears inapplicable for two-qubit gate construction. In this case (two-qubit), only the GA is utilized, the results obtained are sufficiently accurate.
		
	\end{widetext}
	
	\pagebreak
	
	\bibliography{bibliography}
	
\end{document}